\documentclass[twocolumn]{aastex62}
\usepackage{hyperref}
\hypersetup{bookmarksnumbered=true, bookmarksopen=true}

\usepackage{amssymb,amsmath,amsfonts}
\usepackage{newtxtext, newtxmath}

\allowdisplaybreaks[1]  

\usepackage{savesym}
\savesymbol{tablenum}
\usepackage{siunitx}
\restoresymbol{SIX}{tablenum}

\usepackage{bm}
\usepackage{booktabs}
\usepackage{cleveref}  
\usepackage{paralist}  

\graphicspath{{figs/}}

\newcommand{\refsec}[1]{Section~\ref{#1}}
\newcommand{\reffig}[1]{Figure~\ref{#1}}
\newcommand{\reftab}[1]{Table~\ref{#1}}
\newcommand{\refeqn}[1]{Equation~(\ref{#1})}

\newcommand{\rapo}{r_{\mathrm{apo}}}
\newcommand{\rperi}{r_{\mathrm{peri}}}

\renewcommand{\vr}{v_{\mathrm{r}}}
\newcommand{\vt}{v_{\mathrm{t}}}

\newcommand{\vtot}{{v_{\mathrm{tot}}}}

\newcommand{\rs}{r_{\mathrm{s}}}
\newcommand{\vs}{v_{\mathrm{s}}}

\newcommand{\rhos}{\rho_{\mathrm{s}}}

\newcommand{\rlim}{r_{\mathrm{lim}}}

\newcommand{\msun}{M_{\odot}}

\newcommand{\Mpc}{\mathrm{Mpc}}
\newcommand{\kpc}{\mathrm{kpc}}

\shorttitle{halo mass from satellite phase-space distribution}
\shortauthors{Li et al.}

\begin{document}

\title{\large\textbf{A Versatile and Accurate Method for Halo Mass Determination from Phase-Space Distribution of Satellite Galaxies}}

\author[0000-0001-7890-4964]{Zhao-Zhou Li}
\altaffiliation{lizz.astro@gmail.com}
\affil{Department of Astronomy, School of Physics and Astronomy,
  Shanghai Jiao Tong University, 800 Dongchuan Road, Shanghai 200240, China}

\author[0000-0002-3146-2668]{Yong-Zhong Qian}
\affil{School of Physics and Astronomy, University of Minnesota, Minneapolis, MN 55455, USA}
\affil{Tsung-Dao Lee Institute, Shanghai Jiao Tong University, 800 Dongchuan Road, Shanghai 200240, China}

\author[0000-0002-8010-6715]{Jiaxin Han}
\altaffiliation{jiaxin.han@sjtu.edu.cn}
\affil{Department of Astronomy, School of Physics and Astronomy,
  Shanghai Jiao Tong University, 800 Dongchuan Road, Shanghai 200240, China}
\affil{Kavli IPMU (WPI), UTIAS, The University of Tokyo, Kashiwa, Chiba 277-8583, Japan}

\author{Wenting Wang}
\affil{Kavli IPMU (WPI), UTIAS, The University of Tokyo, Kashiwa, Chiba 277-8583, Japan}
\affil{Department of Astronomy, School of Physics and Astronomy,
  Shanghai Jiao Tong University, 800 Dongchuan Road, Shanghai 200240, China}

\author[0000-0002-4534-3125]{Y. P. Jing}
\altaffiliation{{ypjing@sjtu.edu.cn}}
\affil{Department of Astronomy, School of Physics and Astronomy,
  Shanghai Jiao Tong University, 800 Dongchuan Road, Shanghai 200240, China}
\affil{Tsung-Dao Lee Institute, Shanghai Jiao Tong University, 800 Dongchuan Road, Shanghai 200240, China}

\begin{abstract}

We propose a versatile and accurate method to estimate the halo mass and concentration
from the kinematics of satellite galaxies. We construct the 6D phase-space distribution 
function of satellites from a cosmological simulation based on the similarity of internal
dynamics for different halos. Within the Bayesian statistical framework, not only can we 
infer the halo mass and concentration efficiently, but also treat various observational 
effects, including the selection function, incomplete data, and measurement errors,
in a rigorous and straightforward manner. Through tests with mock samples, we show that 
our method is valid and accurate, and more precise than pure steady-state methods. It can constrain the halo mass to within $\sim 20\%$ using only 20 tracers and 
has a small intrinsic uncertainty of $\sim 10\%$. In addition to the clear application
to the Milky Way and similar galaxies, our method can be extended to galaxy groups or clusters.

\end{abstract}


\section{Introduction} \label{sec:intro}

In this paper we present a method to determine the mass of a dark matter halo
using its satellite systems as dynamical tracers. We focus on halos of galactic scale.
The halo mass of a galaxy plays a critical role in connecting observations
to theoretical understandings based on the underlying matter distribution (see e.g., \citealt{Wechsler2018}),
and the halo mass of our own Milky Way (MW) is of particular interest.
Various methods have been applied to estimate the halo mass 
(see \citealt{Courteau2014} and \citealt{Pratt2019} for general reviews, 
and \citealt{Wang2015b}; Wang et al.\ 2019 in preparation for a comprehensive summary of recent estimates for the MW).
These methods are suitable for different scopes and probe different aspects of the mass distribution.
Among them, dynamical modeling with tracers is perhaps one of the most direct approaches.
The mass distribution of the
inner halo can be probed by the kinematics of such tracers as stars, 
planetary nebulae, and globular clusters.
All of the above tracers, 
however, are generally limited in distance from the halo center.
In addition, it was shown that stars could be 
stochastically-biased tracers for the total halo mass
due to incomplete phase-mixing
(e.g., \citealt{Han2016a,Wang2017b,Wang2018}).
Because satellite galaxies have an extended spatial distribution
and can be more easily observed out to large distances by current surveys,
they are expected to be better tracers for the outer halo, 
especially if one is interested in measuring the total halo mass (see \citet{Han2019} for more detailed discussions).

Previous approaches to dynamical modeling with satellite systems include
the virial theorem \citep[e.g.,][]{Biviano2006,Saro2013,Tempel2014a},
the caustic method \citep[e.g.,][]{Diaferio1997,Serra2011,Gifford2013},
the Jeans equation \citep[e.g.,][]{Lokas2002,Evans2003,Watkins2010a,More2011,Mamon2013},
and the phase-space distribution function (DF) \citep[e.g.,][]{Wilkinson1999,Wojtak2009}.
Readers are referred to \citet{Old2015} and \citet{Armitage2019} for systematic 
comparisons of these approaches and to e.g., \citet{Watkins2010a} and \citet{Eadie2015}
for application of the latter two methods to estimate the MW halo mass.
In view of the limited number of satellite galaxies, the associated substantial
statistical uncertainties, and the observational errors,
it is reasonable to make appropriate assumptions to simplify the modeling,
even at the risk of introducing systematic uncertainties.
Several or all of the following assumptions are
commonly made for the above methods: dynamical equilibrium, spherical symmetry, 
a velocity anisotropy profile, and a specific form of the DF.

A nearly minimal but very effective assumption is
that satellite galaxies are in approximate dynamical equilibrium with their host halo,
and therefore, their kinematics can be described by a steady-state phase-space DF. 
While this assumption is supported by 
cosmological simulations, understanding the pertinent DF is a long-standing problem
\citep{Lynden-Bell1967}. Despite many analytical and empirical (simulation-based) 
attempts (e.g., \citealt{Wilkinson1999,Evans2006,Wojtak2008,Williams2015,Posti2015}), an accurate 
and explicit form of the DF for satellite kinematics remains to be found and verified.
As shown by \citet{Wang2015b} and \citet{Han2016a}, introducing unjustified assumptions into the construction of the DF may lead to substantially biased results.
When the 6D phase space coordinates of the tracers are available, 
a promising solution is to use the observed data directly to construct a data-driven DF.
An example is the oPDF (orbital probability DF) method \citep{Han2016b}.
In some other cases, numerical simulations can be used to provide non-parametric templates as empirical DFs.

In a previous study by \citet{Li2017}, the probability density function
in the $(E,L)$ space, $d^2N/(dEdL)=p(E,L)$, where $E$ and $L$ are the energy and 
angular momentum per unit mass, respectively, was derived for satellite galaxies 
directly from cosmological simulations. It was
assumed that the internal dynamics of different halos is similar after the
radius and velocity are normalized by their virial scales. By using the 
$p(E,L)$ for suitable subhalos of a representative template halo, the mass of any other test 
halo can be determined with good accuracy from the kinematic data on a relatively small 
number of satellites (see also \citealt{Callingham2018}). We note, however, that the 
adopted $p(E,L)$ introduces a bias in the halo mass estimate, which can be overcome 
by a simple correction factor. This bias arises because $E$ is not a direct observable. 
As we show in this paper, 
the use of the proper DF, $d^6N/(d^3{\bm{r}}d^3{\bm{v}})=f({\bm{r}},{\bm{v}})$, 
which is in terms of the direct observables (position and velocity vectors, ${\bm{r}}$
and ${\bm{v}}$, respectively) and
related to $p(E,L)$ by transformation of variables, gives an unbiased estimate of the
halo mass, $M$.

The improved method presented here builds upon the previous work of
\citet{Li2017}, but uses the proper DF. It can be understood as a combination of the template-based method of \citet{Li2017} and the oPDF method of \citet{Han2016b}. The new method assumes a Navarro-Frenk-White (NFW, \citealt{Navarro1996}) density profile for the total mass distribution in a halo. 
The universality of this profile supports the similarity of internal dynamics of halos.
Further, the parameters characterizing the NFW profile provide the natural scales for 
normalizing the ${\bm{r}}$ and ${\bm{v}}$ of satellites so that the above similarity can be 
exhibited. Using this similarity, we generalize the simulation-based DF to halos with
any set of $(M,c)$, where $c$ is the concentration parameter for the NFW profile. 
Consequently, this DF provides estimates of both the halo mass and concentration, 
from which the mass distribution can be obtained. In addition, it facilitates a proper 
and straightforward treatment of various observational effects, including selection 
functions, incomplete measurement (e.g., lack of proper motion),
and observational errors.

Due to diversities in formation history and environment, individual halos are 
expected to exhibit deviations from our assumed DF. This 
halo-to-halo scatter represents the systematic uncertainty introduced by our 
assumptions. Using a large mock sample of realistic halos from a cosmological simulation,
we demonstrate the validity of our method and quantify its systematic uncertainty.
While the major motivation for this work is to 
estimate the mass of the MW halo with better precision
(see e.g., Wang et al.\ 2019 in preparation for recent estimates and uncertainties),
the method in principle can be extended to any halos, including galaxy clusters.

The layout of this paper is as follows. We outline our method in \refsec{sec:method}
and show how to construct the simulation-based DF of satellite kinematics in 
\refsec{sec:phasespace}. We test the validity and precision of our method using
mock samples and make comparisons with other methods in \refsec{sec:mocktest}.
We discuss applications of our method in \refsec{sec:discussion}.
Further discussion and conclusions are given in \refsec{sec:conclusion}.

\section{Outline of the Method}\label{sec:method}

As is true of dynamical modeling in general, an accurate statistical description of 
the satellite kinematics is essential to our method of estimating the halo mass.
Our description is based on the following assumptions:

(1) All halos have the spherical NFW density profile
\begin{equation}
\rho(r)=\frac{\rhos}{(r/\rs)(1+r/\rs)^2},
\end{equation}
where $r$ is the radius, and $\rhos$ and $\rs$ are the characteristic scales for density 
and radius, respectively;

(2) The satellites are in dynamical equilibrium with its host halo and their kinematics 
in terms of the radial vector, $\bm{r}$, and velocity, $\bm{v}$, can be described by a 
steady-state DF in phase space
\begin{equation}
\frac{d^6N}{d^3{\bm{r}}d^3{\bm{v}}}=f({\bm{r}},{\bm{v}}),
\label{eqn:frv}
\end{equation}
which is normalized as $\int f({\bm{r}},{\bm{v}})d^3{\bm{r}}d^3{\bm{v}}=1$;

(3) The internal dynamics of all halos are similar after $\bm{r}$ and 
$\bm{v}$ are normalized by their characteristic scales, $\rs$ and 
$\vs=\rs\sqrt{4\pi G\rhos}$\,, 
respectively, where $G$ is the gravitational constant.

For a spherical system with a stationary potential, the orbit of a tracer
can be specified by the energy per unit mass,
\begin{equation}
  E = \frac{1}{2} (\vr^2+\vt^2) + \Phi (r),
  \label{eqn:ephi}
\end{equation}
and the angular momentum per unit mass, $ L = r \vt$, of the tracer,
where $\vr$ and $\vt$ are the radial and tangential components of its velocity,
respectively.
In \refeqn{eqn:ephi},
\begin{equation}
\Phi(r) = \vs^2 \left[ 1 - \frac{\ln (1 + r / \rs)}{r / \rs} \right],
\end{equation}
with $\Phi(0)=0$, is the gravitational potential\footnote{
Our choice of $\Phi(0)=0$ gives a more consistent definition of the potential,
which makes it easier to compare or combine the DFs for distinct isolated halos. 
For example, two isolated halos with identical internal density profiles 
but surrounded by different spherical mass distributions have identical
internal dynamics, for which it is more appropriate to use identical potentials. 
Our choice gives the same potential for both halos, but the common choice of 
$\Phi(\infty)=0$ does not.}
corresponding to the NFW profile.

Because only $({\bm{r}},{\bm{v}})$ can be obtained directly from observations, 
the probability density function in the 6D phase space of $({\bm{r}},{\bm{v}})$,
${d^6N}{/(d^3{\bm{r}}d^3{\bm{v}})}=f({\bm{r}},{\bm{v}})$, 
instead of that in the 2D $(E,L)$ space, ${d^2N}{/(dEdL)}=p(E,L)$, 
is the proper DF to use in a maximum likelihood estimate 
of the halo properties from the kinematics of satellites.
We will construct $f({\bm{r}},{\bm{v}})$ from a cosmological simulation. Due to the limited number of 
satellites for MW-like halos in this simulation, however, it is difficult to obtain an
accurate $f({\bm{r}},{\bm{v}})$ directly from the representation of satellites in the 6D phase space of 
$({\bm{r}},{\bm{v}})$. It is more practical to first construct $p(E,L)$ from the representation 
in the 2D $(E,L)$ space based on the simulation, and
then use \refeqn{eqn:fel} below to convert $p(E,L)$ into $f({\bm{r}},{\bm{v}})$.
For clarity and convenience, we will refer to $f({\bm{r}},{\bm{v}})$ as the DF and $p(E,L)$ as the orbital distribution.

Assuming spherical symmetry,
we can decompose the DF $f({\bm{r}},{\bm{v}})$ into
the orbital distribution $p(E,L)$ and the radial distribution $p(r|E,L)$ along the orbit
with a specific set of $(E,L)$,
\begin{equation}
f({\bm{r}},{\bm{v}})=\frac{|\vr|}{8\pi^2L}p(r|E,L)p(E,L),
\label{eqn:frvprel2}
\end{equation}
where the first term on the right-hand side comes from the Jacobian for transformation of variables
(see Appendix~\ref{sec:felpel} for details).
Under the steady-state assumption (see e.g., \citealt{Han2016b}), 
\begin{equation}
p(r|E,L)dr=\frac{dr}{|\vr(r,E,L)|T_{\mathrm{r}}(E,L)},
\label{eqn:prel}
\end{equation}
where $\vr(r,E,L)={\pm\sqrt{2[E-\Phi(r)]-L^2/r^2}}$ is the radial velocity,
\begin{equation}
T_{\mathrm{r}}(E,L)=2\int_{\rperi}^{\rapo}\frac{dr}{|\vr(r,E,L)|}
\label{eqn:tr}
\end{equation}
is the radial orbital period for a specific set of $(E,L)$, and $\rperi$ and $\rapo$ are 
the radii at the pericenter and apocenter, respectively.
So we obtain
\begin{equation}
f({\bm{r}},{\bm{v}})=\frac{p(E,L)}{8\pi^2LT_{\mathrm{r}}(E,L)}.
\label{eqn:fel}
\end{equation}

As we can see, the DF $f({\bm{r}},{\bm{v}})$ can be fully specified in terms of two isolating integrals, $E$ and $L$,
in accordance with Jeans theorem (section 4.2 of \citealt{Binney2008}). Mathematically, $f({\bm{r}},{\bm{v}})$ only
depends on $(r,\vr\,,\vt)$ under our assumption of spherical symmetry. So we can write $f({\bm{r}},{\bm{v}})=f(r,\vr\,,\vt)$
by dropping the irrelevant variables. On the other hand, $E$ is a function of $(r,\vr\,,\vt)$ and $L$ is a function of $(r,\vt)$. Therefore, Jeans theorem dictates that the dependence of $f(r,\vr\,,\vt)$ on $(r,\vr\,,\vt)$ can only be in
terms of $E(r,\vr\,,\vt)$ and $L(r,\vt)$. With this understanding, we write, for convenience, 
\begin{equation}
f({\bm{r}},{\bm{v}})=f(r,\vr\,,\vt)=f(E,L).
\end{equation}
Because the DF $f({\bm{r}},{\bm{v}})$ contains the information on both the orbital distribution $p(E, L)$ and the radial distribution $p(r|E,L)$ for each orbit, we expect that the DF method may constrain halo properties better than other methods based on less information (see \refsec{sec:comparison}).

Under our assumption (3), both $p(E,L)$ and $f(E,L)$ have similar forms for all halos 
after $\bm{r}$ and $\bm{v}$ are normalized by the characteristic scales $\rs$ and
$\vs$, respectively. In other words, using  the dimensionless variables 
$\tilde r=r/\rs$, $\tilde v_{\mathrm{r}}=\vr/\vs$, $\tilde v_{\mathrm{t}}=\vt/\vs$,
and others derived from them, such as $\tilde E=E/\vs^2$ and $\tilde L=L/(\rs\vs)$,
we can obtain the dimensionless $\tilde p(\tilde E,\tilde L)$ and 
$\tilde f(\tilde E,\tilde L)$ that are universal to all halos. The normalization
$\int\tilde f(\tilde E,\tilde L)d^3\tilde{\bm{r}}d^3\tilde{\bm{v}}=\int f(E,L)d^3\bm{r}d^3\bm{v}=1$
requires
\begin{equation}
f(E,L)=\frac{1}{\rs^3 \vs^3} \tilde{f}(\tilde E, \tilde L).
\end{equation}
For a halo corresponding to the NFW profile with characteristic scales $\rs$ and $\vs$,
we define its mass, $M$, and the associated virial radius, $R$, so that the mean density 
within $R$ is 200 times the critical density of the present universe. It is straightforward
to show that 
\begin{equation}
  \vs^2 = \frac{GM}{R} \left[\frac{\ln (1 + c)}{c} - \frac{1}{1 + c}\right]^{-1},
\end{equation}
where $c=R/\rs$ is the concentration of the halo. Therefore, $M$ and $c$ are equivalent 
to $\rs$ and $\vs$. To emphasize that the halo properties are estimated from the raw 
kinematic data through the proper DF discussed here, we denote this essential input to 
our method as
\begin{equation}
  p (\bm{w}|M, c) = f(E,L),
\end{equation}
where $\bm{w}$ is a shorthand for $(\bm{r}, \bm{v})$.

\subsection{Estimate of halo properties}

Consider a sample of $N_\mathrm{tracer}$ satellites with mutually independent kinematic
data $\{ {{\bm{w}}}_i\,, i = 1, \cdots, N_\mathrm{tracer}\}$. Using the DF discussed 
above, we can write the probability of observing such a sample for a halo of mass, $M$, 
and concentration, $c$, as
\begin{gather}
  p (\{ {{\bm{w}}}_i \} |M, c) = \prod^{N_\mathrm{tracer}}_{i = 1} 
  p ({{\bm{w}}}_i |M, c).
  \label{eqn:likelihood}
\end{gather}
The best-fit values of $(M,c)$ can then be inferred from the maximum likelihood method
with the Bayesian formula
\begin{gather}
  p (M, c| \{ {{\bm{w}}}_i \}) \propto p (\{
  {{\bm{w}}}_i \} |M, c) p (c|M) p (M), \label{eqn:posterior}
\end{gather}
where $p (M)$ and $p (c|M)$ represent our prior knowledge of the halo.
For a random halo in the universe, the halo mass function
and the mass-concentration relation are the natural choices for these two priors.
One can also use a flat prior to avoid relying on extra information.

\subsection{Inclusion of observational effects} \label{sec:obser_effects}

Real data suffer from various observational effects, such as incompleteness
due to sample selection, lack of certain measurements (e.g., proper motion),
and observational errors. It is straightforward to include these effects in
deriving the proper DF. We give two types of examples below.

\subsubsection{Selection function}\label{sec:select}

For a sample restricted by the selection function $S({\bm{w}})$,
the DF is given by
\begin{equation}
  p_{\mathrm{s}} ({\bm{w}}|M, c) = 
  \frac{p ({\bm{w}}|M, c) S ({\bm{w}})}
       {\int p ({\bm{w}}' |M, c) S ({\bm{w}}') d^6{\bm{w}}^{\prime}},
\end{equation}
where $d^6{\bm{w}'}$ is a shorthand for $d^3\bm{r}'d^3\bm{v}'$.
The selection function $S({\bm{w}})\leqslant 1$ indicates the degree
of completeness by specifying how likely a satellite with the
kinematic data $\bm{w}$ is observed. A simple example is
a sample that is complete within some distance.
More realistically, because observations of satellites are restricted by 
the limiting magnitude of the survey (e.g., \citealt{Jethwa2018}),
less luminous satellites are observed out to smaller distances.

\subsubsection{Incomplete data} \label{sec:incomp_info}

It is very difficult to measure the tangential velocity (proper motion)
for distant satellites of the MW, not to mention satellites of extragalactic 
systems. The radial velocity might also be absent, as its measurement
requires high-quality spectrum, and hence, expensive time on large 
telescopes. To maximize the use of all available data and improve the 
estimates, we can include satellites with incomplete data
by marginalizing the DF. For those without $\vt$ or $\vr$, 
the marginalized DF is
\begin{align}
  p (\bm{r}, \vr|M,c)& = 2 \pi \int f (r, \vr\,, \vt) \vt d \vt\,,
  \label{eqn:vt}\\
  p ({\bm{r}}, \bm{v}_\mathrm{t}|M,c)& = \int f(r, \vr\,, \vt) d \vr\,,
  \label{eqn:vr}
\end{align}
respectively. Note for example, that $p (\bm{r}, \vr|M,c)$ is evaluated at
fixed $(r,\vr)$, which, along with the integration variable $\vt$, determine
the input $(E,L)$ for $f (r, \vr\,, \vt)=f(E,L)$.

From a theoretical perspective, previous studies 
\citep[e.g.,][]{Binney2008,Wojtak2008} showed that the mass distribution 
of a spherical system is mainly determined by the energy distribution of its
constituents, but is insensitive to their velocity anisotropy or angular momentum 
distribution. Therefore, we may obtain
good estimates of the halo properties by using the marginalized DF for
the total velocity $\vtot=\sqrt{\vr^2+\vt^2}$\,, which is given by
\begin{align}
  p ({\bm{r}}, \vtot|M,c) &= 2 \pi\int \delta \Big(\vtot - \sqrt{\vr^2 + \vt^2}\Big)
    f (r, \vr\,, \vt)  \vt d \vt d \vr\nonumber\\
   &= 2 \pi \vtot \int_{- \vtot}^\vtot f\Big(r, \vr\,, \sqrt{\vtot^2-\vr^2}\Big) d \vr.
  \label{eqn:vv}
\end{align}

The above results are applied in \refsec{sec:info_used} to discuss the
effects of individual observables on the estimates of halo properties.

\section{Construction of phase-space distribution}\label{sec:phasespace}

We now present the details of constructing the DF of satellites in the
phase space of $(\bm{r},\bm{v})$ based on a cosmological simulation.

\subsection{Halo sample} \label{sec:simulation}

We use a large sample of halos selected from 
a cosmological N-body (dark matter only) simulation,
the Millennium II \citep{Boylan-Kolchin2009}.
A similar sample was used to study the systematics
of dynamical modeling in \citet{Wang2017b,Wang2018}. 
The Millennium II simulation has a box size of $100 h^{- 1} {\mathrm{Mpc}}$
and a particle mass of $m_p = 6.9 \times 10^6 h^{- 1} M_{\odot}$.
It adopts the first-year WMAP cosmology \citep{Spergel2003} with
$\Omega_m = 0.25$, $\Omega_{\Lambda} = 0.75$, $h = 0.73$, $n_s = 1$, 
and $\sigma_8 = 0.9$.

We identify a sample of halos with
$11.5 \leq \lg\ M/M_\odot \leq 12.5$ that are analogous to the MW in 
mass.\footnote{
 For ease of comparison with results in the literature, we define the halo mass,
 $M$, and the associated virial radius, $R$, so that the mean density within
 $R$ is 200 times the critical density of the present universe.}
 As shown in \citet{Li2017}, the influence of a massive neighbor (e.g., M31)
on satellite kinematics is small when the neighbor is at a distance exceeding 
3 times its virial radius. To exclude potential influence of nearby massive
neighbors, we select isolated halos by requiring that 
all companions within a sphere of $2\ \Mpc$ in radius are at least an order of 
magnitude smaller in mass. In the end, we have 943 isolated halos, each of 
which contains $\sim 10^5$ particles.

\begin{figure}[tbp]
  \centering
  \includegraphics[width=0.45\textwidth]{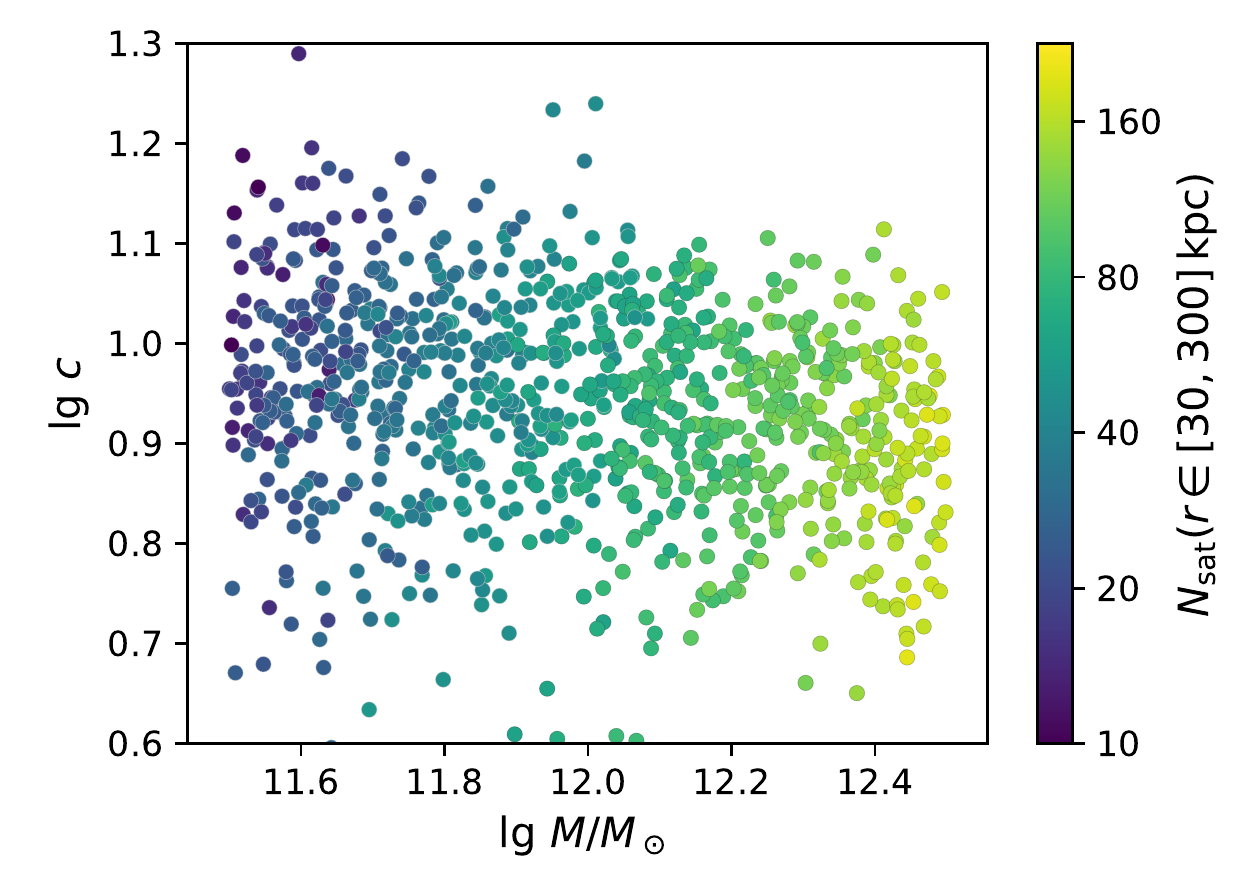}
  \caption{
   Fitted values of the halo mass, $M$, and concentration, $c$, for our halo sample
   based on the NFW profile. Each dot represents a halo and is colored according to 
   the number of satellites in the radial range of $r=30$--300~$\kpc$ from the halo center.
  }
  \label{fig:Mh_vs_c}
\end{figure}

We fit the NFW density profile to each halo in the sample.
The fitted values of the halo mass, $M$, and concentration, $c$, are shown
in \reffig{fig:Mh_vs_c}. The NFW profile is a good fit in general, with 
differences of $\lesssim 3\%$ between the fitted $M$ and the true values.
We drop 3 halos with uncommonly small values of $c<3.5$.
The concentration parameters of the remaining 940 halos follow a log-normal 
distribution with a slight dependence on the halo mass. 
We take this result as the default prior for $c$:
\begin{equation}
p(\lg c|M)=\mathcal{N} (0.94 - 0.077 \lg (M / 10^{12} M_{\odot}), 0.11).
\label{eqn:lgc}
\end{equation}
The above result is consistent with previous studies (e.g., \citealt{Dutton2014}).
The halos in the final sample have characteristic radial scales
of $r_s\sim 20$--$40\ \kpc$. As discussed in \refsec{sec:method}, we will
use the $\rs$ and $\vs=\rs\sqrt{4 \pi G \rhos}$ associated with the NFW profile 
to obtain dimensionless variables such as $\tilde r=r/\rs$ and $\tilde\vr=\vr/\vs$.

\subsection{Satellite samples}\label{sec:satellite}

Based on the Millennium II simulation, \cite{Guo2011} generated
a galaxy catalog using the semi-analytical model (SAM) for galaxy formation.
We select satellites from this catalog, which includes
orphan satellites whose parent subhalos are eventually disrupted. 
We only use satellites with stellar masses of $m_\star \geq 100\,\msun$ 
whose parent subhalos have infall masses of $M_\mathrm{inf} \geq 30\,m_p$.

\begin{figure}[tbp]
  \centering
  \includegraphics[width=0.45\textwidth]{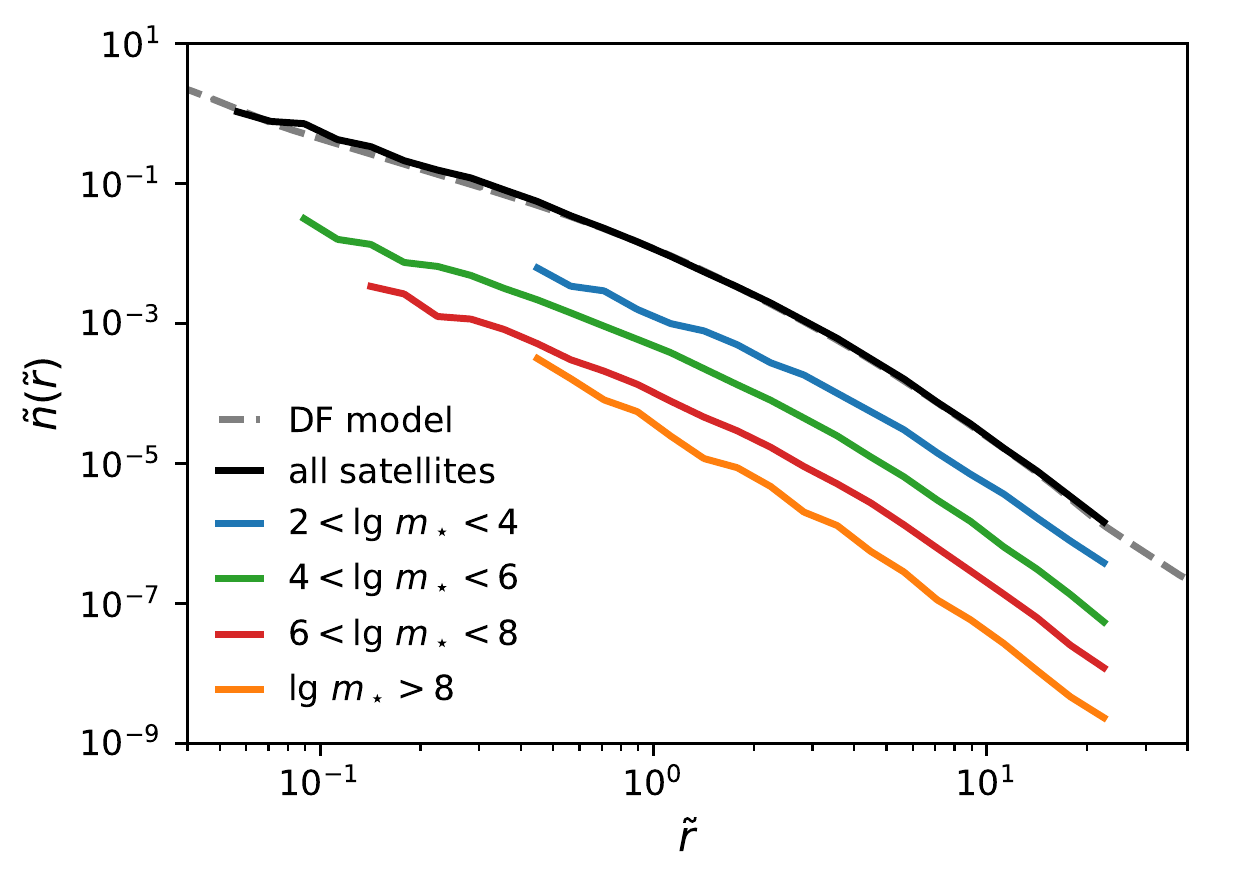}
  \caption{
     The spatial distributions of satellites with $\tilde r=r/\rs \leq 25$
     in the template sample.
     The black solid curve shows the normalized dimensionless number density
     $\tilde n$ as a function of $\tilde r$ for all of the selected satellites.
     This result is reproduced very well by the dashed gray curve derived from
     the simulation-based DF.
     The colored curves show the distributions for satellites with different stellar 
     masses. These curves are shifted in amplitude for better comparison.
     The true distributions are truncated at the inner radius bin that contains
     fewer than 10 satellites. They show little dependence on stellar masses
     of satellites for $\tilde r>1$.
  }
  \label{fig:space_distr}
\end{figure}

We select a sample of 104,315 satellites with radii $r \leq 25 \rs$ (measured from
the respective host halo centers), hereafter referred to as the template sample,
to construct the DF in the phase space of
$(\bm{r},\bm{v})$. The Hubble flow is included when we calculate the
$\bm{v}$ for satellites relative to their host halos, but this inclusion only
makes a small difference. Combining the statistics on satellites with the same 
dimensionless $\tilde r=r/\rs$, we show their spatial distributions in 
\reffig{fig:space_distr}.
It can be seen that these distributions depend very little on stellar
masses of satellites, at least for $\tilde r >1$. Although not shown in 
\reffig{fig:space_distr}, the spatial distributions of satellites  for $\tilde r>1$ 
do not depend on infall masses of their parent subhalos, either. 
These results are consistent with those in literature 
(e.g., \citealt{Han2016,Newton2018}) and suggest that
selection based on satellite luminosity or mass may not introduce any
significant bias.

To mimic observations of the MW satellites, we make mock samples of
satellites for individual halos by selecting only those satellites with 
$30 \leq r \leq 300\ \kpc$. The number of such satellites for each halo ranges from 
$\sim 10$ to 200, and is shown by the color coding in \reffig{fig:Mh_vs_c}.

\subsection{Phase-space distribution} \label{sec:phase_distr}

Following our assumptions in \refsec{sec:method}, especially the one regarding 
the similarity of internal dynamics for different halos, we construct the dimensionless 
DF in phase space, $\tilde{f} (\tilde{E}, \tilde{L})$, by combining the statistics on 
the satellites in the template sample. 
The validity of our assumptions is discussed in Appendix~\ref{sec:equilibrium} and
ultimately tested by applying the above DF to 
estimate halo properties of mock samples in \refsec{sec:mocktest}.
Details of constructing the DF are given in Appendix~\ref{sec:smooth}.
A similar procedure was used in \citet{Callingham2018}, 
but to construct $\tilde{p} (\tilde{E}, \tilde{L})$ only.

It is worthwhile to point out some subtlety in constructing $\tilde{f} (\tilde{E}, \tilde{L})$.
As defined in \refeqn{eqn:tr}, the radial orbital period $T_r (E, L)$ used for the construction 
corresponds to $\rperi\leq r\leq\rapo$ in general.
However, because the satellites in the template sample have $r \leq r_{\lim}=25 \rs$, the upper limit for the radial integral in \refeqn{eqn:tr} should be replaced by the smaller
of  $r_{\lim}$ and $\rapo$. Nevertheless, it is important to recognize that the DF
constructed from this sample, $\tilde{f} (\tilde{E}, \tilde{L})$, is a function of
$\tilde{E}$ and $\tilde{L}$ only, and is not subject to any radial limit. In other words,
so long as our assumptions are valid, all satellites with different 
$(\tilde{\bm{r}},\tilde{\bm{v}})$ but the same $(\tilde{E},\tilde{L})$ are described by
$\tilde{f} (\tilde{E}, \tilde{L})$.
(If the steady-state assumption is not satisfied everywhere, 
the obtained DF is essentially the average within $r_{\lim}$ for a halo.)

\begin{figure}[htbp]
  \centering
  \includegraphics[width=0.45\textwidth]{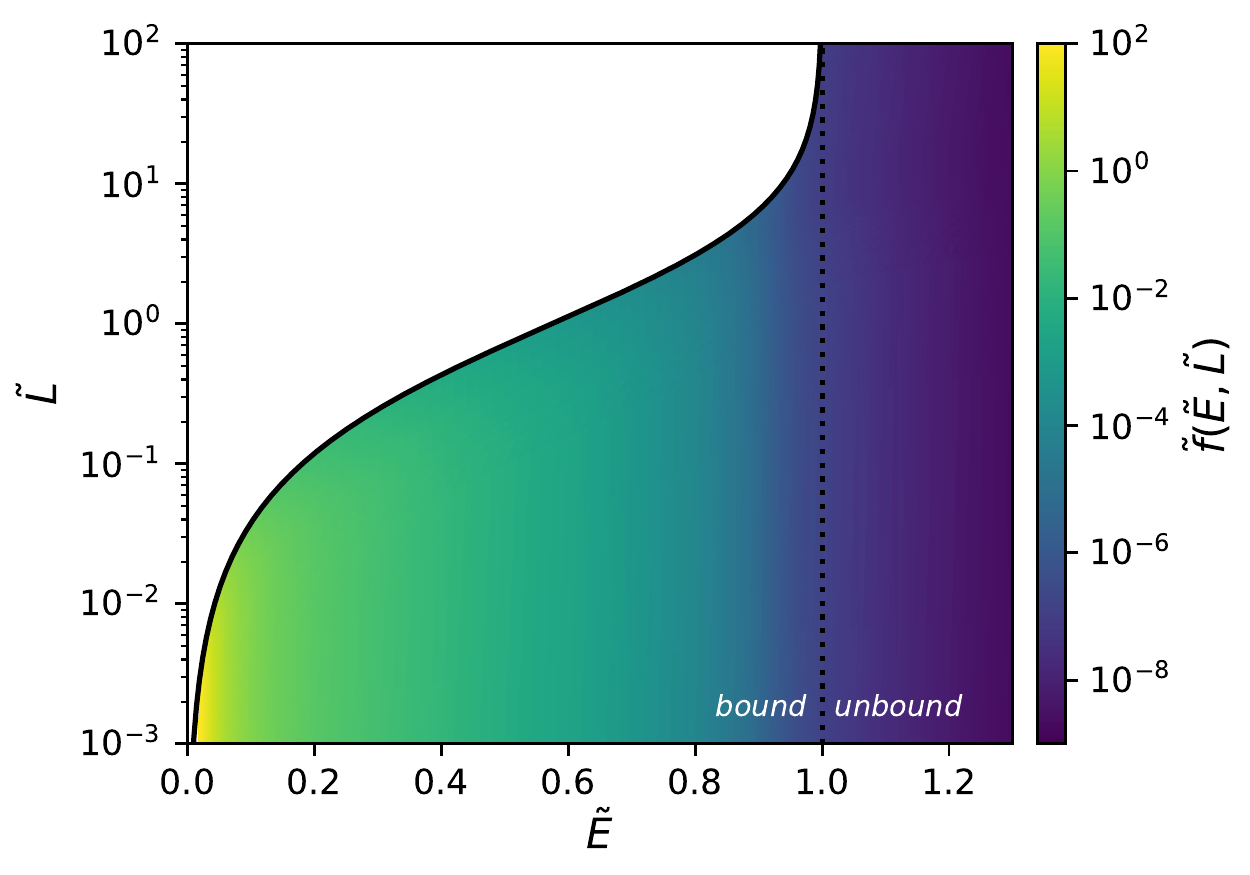}
  \caption{
    Dimensionless DF in phase space, $\tilde{f} (\tilde{E}, \tilde{L})$,
    constructed from the template sample of satellites. Note that satellites with
    $\tilde{E}>1$ (to the right of the dotted line) are not bound.
   }
  \label{fig:DF}
\end{figure}

Our simulation-based DF, $\tilde f(\tilde E, \tilde L)$, is shown in \reffig{fig:DF}.
This DF of satellites is very similar to that obtained by \citet{Wojtak2008} for dark matter 
particles in simulated cluster halos. In both cases, the DF depends largely on $\tilde E$ 
and is not very sensitive to $\tilde L$. Note that our DF covers unbound satellites,\footnote{
All tracers need not be bound to their host halo in a steady state, which only requires them 
to be uniformly distributed in phase angles along the orbits \citep{Han2016b}.} 
which constitute $\approx0.35\%$ of the template sample.
Therefore, in using this DF to estimate the properties of
a halo, no assumption needs to be made regarding whether any
satellite is bound to it or not (e.g., Leo I for the MW).

As a simple test of consistency, we calculate the dimensionless density profile
for the satellites in the template sample,
\begin{equation}
 \tilde n (\tilde r) =2\pi\int\tilde f (\tilde E,\tilde L)\tilde v_{\mathrm{t}}
 d\tilde v_{\mathrm{t}} d\tilde v_{\mathrm{r}}\,,
\end{equation}
where $\vt$ is integrated from 0 to $\infty$ and $\vr$ from $- \infty$ to $\infty$.
As shown in \reffig{fig:space_distr}, the above result is in excellent agreement 
with the true profile.

\section{Validation with Mock Tests} \label{sec:mocktest}

We test the validity and accuracy of our method for estimating the halo mass, $M$,
and concentration, $c$, using the mock samples of satellites described in 
\refsec{sec:satellite}. For each test halo (see \refsec{sec:simulation}),
we estimate its properties by choosing a random subset of its satellites as observed 
tracers. Because the satellites in the mock samples have 
$30\leq r\leq 300\ \kpc$, the proper DF to use (see \refsec{sec:select}) is
\begin{equation}
   p_{\mathrm{s}} ({\bm{w}}|M, c) = 
  \frac{p ({\bm{w}}|M, c)}
       {\int_{30\leq r'\leq 300\ \kpc} p ({\bm{w}'}|M, c)d^6{\bm{w}}'}.
       \label{eqn:dfselect}
\end{equation}
The above DF is used in \refeqn{eqn:posterior} to estimate halo properties with
a flat prior in terms of $\lg M$. The default prior on $c$ is given in \refeqn{eqn:lgc}.
We also present results for a flat prior in terms of $\lg c$.

\subsection{Example application to a halo} \label{sec:single}

\begin{figure}[htbp]
  \centering
  \includegraphics[width=0.45\textwidth]{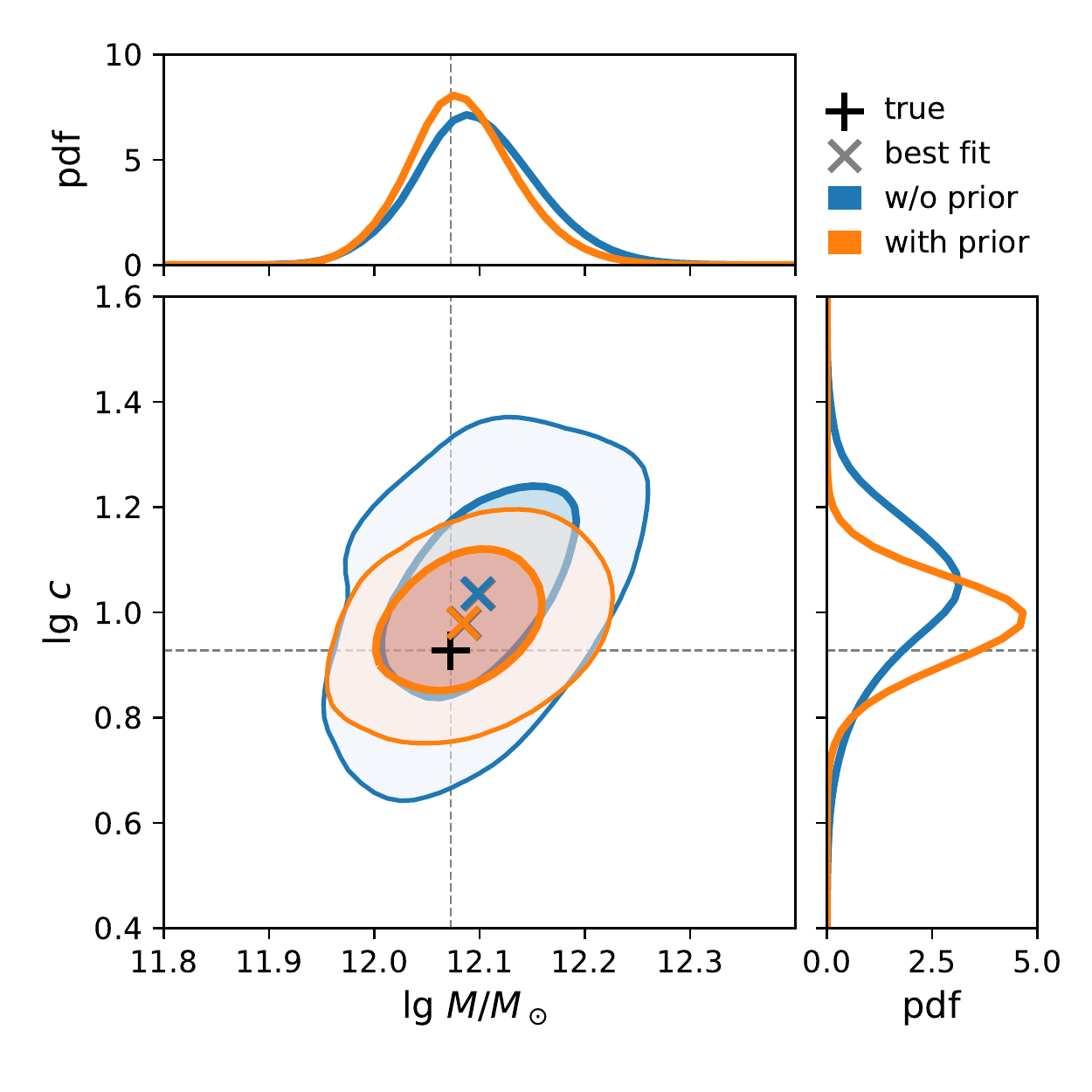}
  \caption{
  Estimated $\lg M$ and $\lg c$ from the kinematics of 40 tracers with $30\leq r\leq 300\ \kpc$
  for a test halo. The results in red and blue are for the default and flat priors on $c$, 
  respectively. The best-fit and true values are indicated by the cross and
  plus symbols, respectively. The dark and faint contours correspond to the $1\sigma$ and 
  $2\sigma$ (68.3\% and 95.4\%) confidence levels. The top and right panels show the 
  marginalized probability distributions for the estimated $\lg M$ and $\lg c$, respectively.
  }
  \label{fig:single_contour}
\end{figure}

We randomly picked a halo from the sample described in \refsec{sec:simulation}, and 
then randomly chose 40 of its satellites with $30\leq r\leq300\ \kpc$ as observed 
tracers. Using the default prior on $c$, we calculate the joint probability distribution 
of $(\lg M,\lg c)$ from \refeqn{eqn:posterior} on a 2D grid. The best-fit
values corresponding to the maximum likelihood, $(\lg M_\mathrm{esti},\lg c_\mathrm{esti})$, 
are indicated by the red cross in \reffig{fig:single_contour}, where the $1\sigma$ and 
$2\sigma$ (68.3\% and 95.4\%) confidence contours are also shown in dark and faint red,
respectively. For comparison, we replace the default prior on $c$ with 
a flat prior and show the corresponding results in blue in \reffig{fig:single_contour}.
From the marginalized distributions of $\lg M$ and $\lg c$ shown in the top and right
panels, respectively, we obtain the statistical uncertainties, 
$\sigma_{\lg M}=0.053$ (0.059) dex and $\sigma_{\lg c}=0.087$ (0.14) dex,
when the default (flat) prior on $c$ is used. The default prior on $c$ 
significantly reduces the uncertainty in the estimated $\lg c$ as expected, but it only 
slightly improves the precision of the estimated $\lg M$.

For both the default and flat priors on $c$, the best-fit value
$\lg M_\mathrm{esti}$ ($\lg c_\mathrm{esti})$ is within $1\sigma_{\lg M}$
($1\sigma_{\lg c}$) of the true value $\lg M_\mathrm{true}$ ($\lg c_\mathrm{true})$.
As shown in \refsec{sec:systematics}, 
the $\sigma_{\lg M}$ and $\sigma_{\lg c}$ obtained in the above example 
application of our method to a single halo are reliable estimates of
the statistical uncertainties, which dominate the total uncertainties when the number
of observed tracers is $<100$.

\begin{figure}[htbp]
  \centering
  \includegraphics[width=0.45\textwidth]{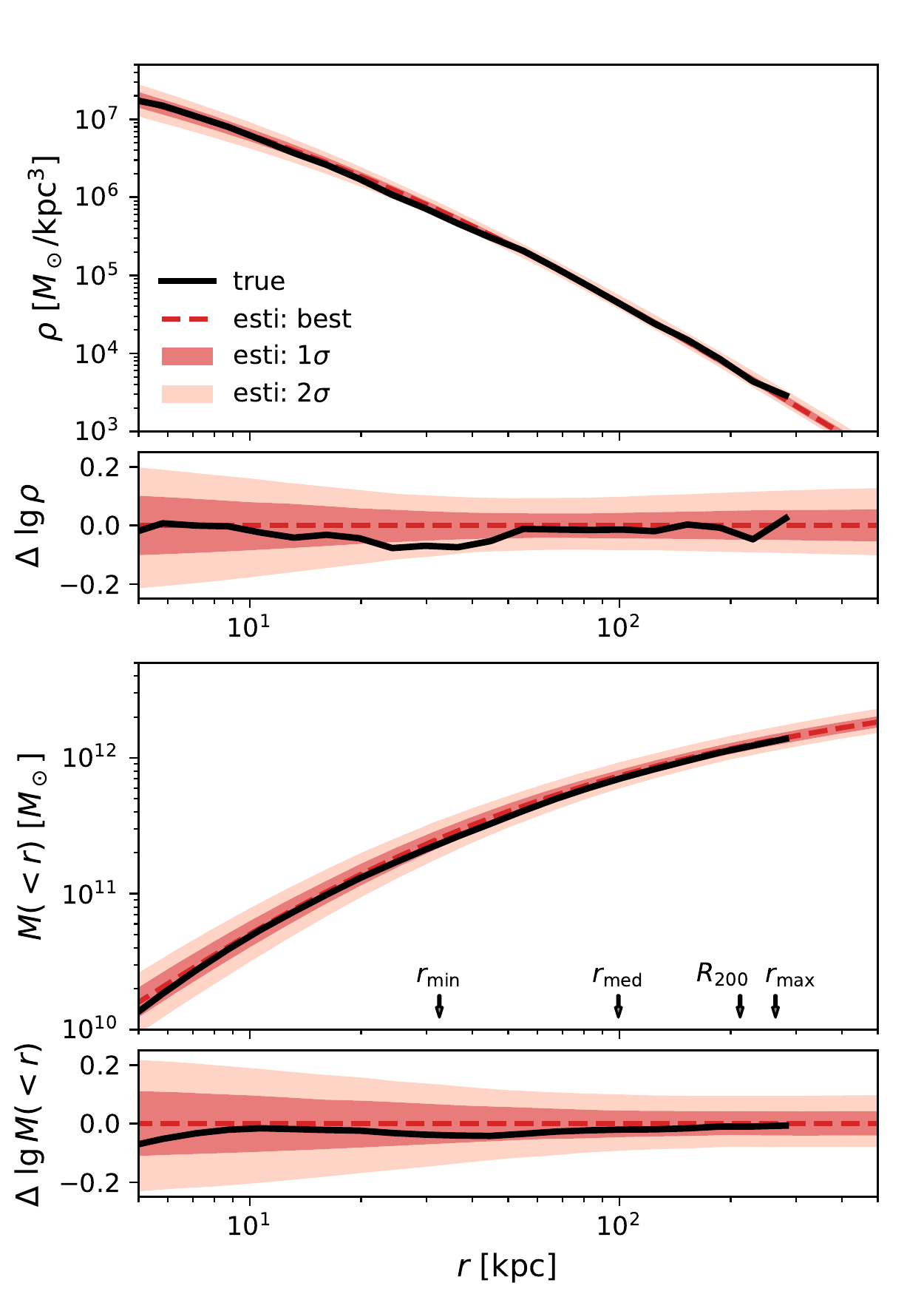}
  \caption{
  Comparison of the estimated $\rho(r)$ and $M(<r)$ (red) with their true counterparts 
  (black) for the same test halo in \reffig{fig:single_contour}. The default prior on
  $c$ is used. The dashed curves and dark (faint) shaded regions correspond to the best 
  estimates and $1\sigma$ ($2\sigma$) confidence regions. The arrows indicate the minimum 
  ($r_\mathrm{min}$), median ($r_\mathrm{med}$), and maximum ($r_\mathrm{max}$) radii 
  of the observed tracers, as well as the virial radius ($R$) of the halo.
  }
  \label{fig:single_compare}
\end{figure}

\reffig{fig:single_compare} shows the estimated density profile, $\rho(r)$, and
mass distribution, $M(<r)$, for the default prior on $c$ and compares them with their 
true counterparts. For either of these estimated functions, deviations from the 
true form are consistent with the estimated confidence levels.
Specifically, $\rho(r)$ is tightly constrained at the $\sim 10\%$ level
for most of the radial range of the observed tracers ($r \gtrsim 30 \kpc$), and $M(<r)$
is best constrained at a similar level beyond the median radius, $r_\mathrm{med}$, 
of the tracers. Similar results are found when the flat prior on $c$ is used
(not shown). Note that our method can provide tight constraints on $\rho(r)$ and $M(<r)$ 
throughout the outer halo. In contrast, the constraints reported by
\citet{Wolf2010a}, \citet{Amorisco2011}, and \citet{Han2016b} quickly deteriorate
away from a certain pinch point (usually near the half-mass radius or 
$r_\mathrm{med}$, see \citealt{Han2016b} for a detailed discussion).

\subsection{General performance} \label{sec:tests}

As shown by the example in \refsec{sec:single}, application of our method to a halo
with a single set of observed tracers not only gives the best-fit values 
$\lg M_\mathrm{esti}$ and $\lg c_\mathrm{esti}$, but also the associated
statistical uncertainties $\sigma_{\lg M}$ and $\sigma_{\lg c}$.
A single application, however, cannot determine the systematic uncertainties. 
In this subsection,
we estimate the total (systematic and statistical) uncertainties using the halo
sample described in \refsec{sec:simulation}. We extend the analyses to separate 
the systematic from the statistical uncertainties in \refsec{sec:systematics}.

We apply our method to three sets of test halos, which differ in the
number of tracers chosen for mock observations per halo. Specifically, 
the number of tracers used is $N_\mathrm{tracer}=10$, 40, or 80, respectively. 
Clearly, for every halo in each set, the number of 
its satellites with $30\leq r\leq 300\ \kpc$ must exceed the corresponding
$N_\mathrm{tracer}$. As demonstrated in Appendix~\ref{sec:robust}, selection
based on the richness of satellites for a halo does not introduce any bias.
We apply our method to each test halo multiple times by randomly choosing 
different tracer subsamples. We include all the estimates obtained 
from this procedure so that a statistical result for each set of test halos is 
based on $\sim 10^4$ estimates.

\begin{figure*}[htbp]
  \centering
  \includegraphics[width=1\textwidth]{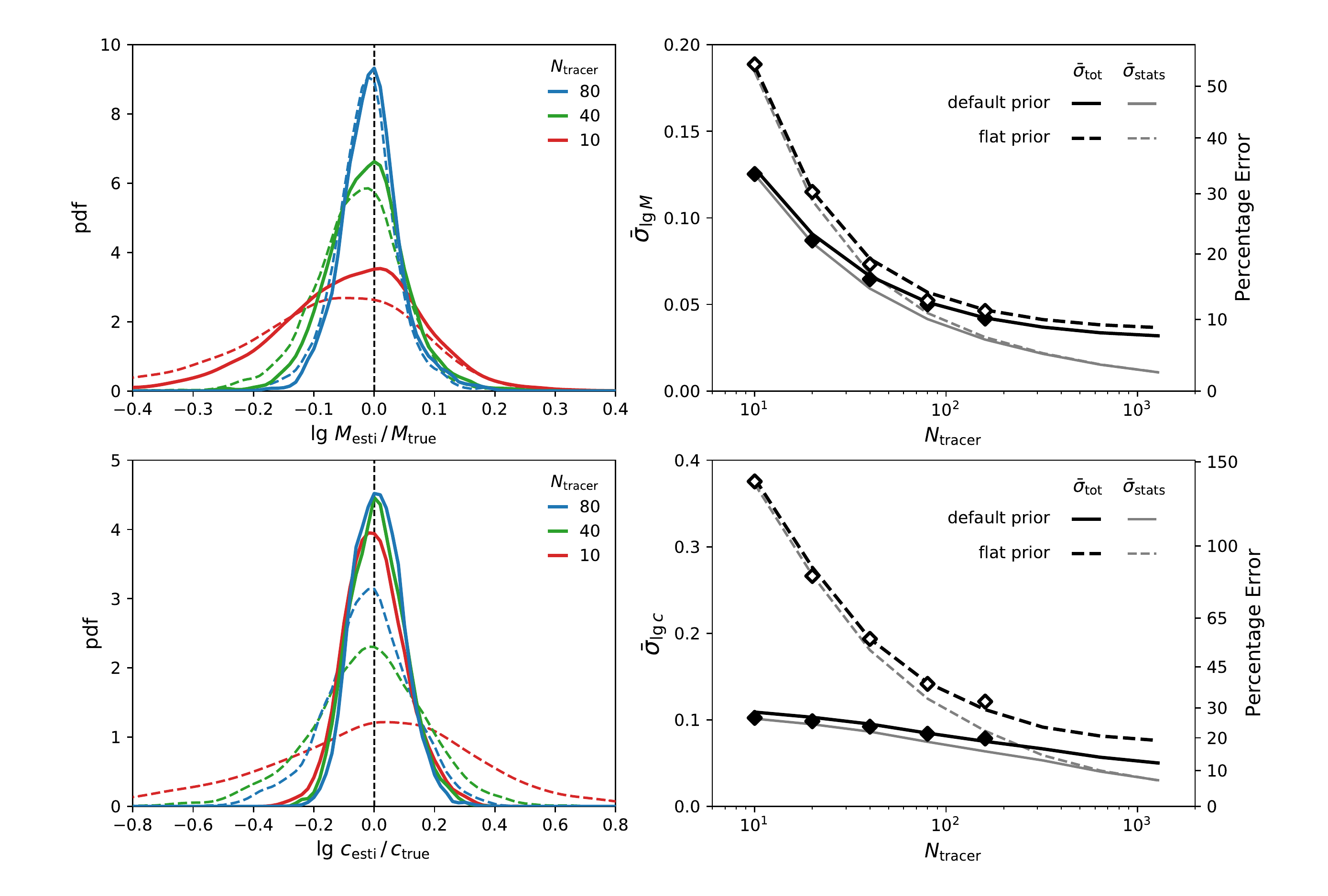}
  \caption{
  Precision of estimated $\lg M$ (upper panels) and $\lg c$ (lower panels)
  for the test halo sample. \textit{Left panels}: Kernel smoothed distributions for the
  deviations of the best-fit values, $\lg M_\mathrm{esti}$ and $\lg c_\mathrm{esti}$, 
  from the true values, $\lg M_\mathrm{true}$ and $\lg c_\mathrm{true}$, respectively.
  Results are shown for three sets of test halos, with a total of $N_\mathrm{tracer}=10$, 
  40, or 80 tracers, respectively, chosen for mock observations per halo. The solid (dashed)
  curves are for the default (flat) prior on $c$. \textit{Right panels}: Uncertainties in
  the estimated $\lg M$ and $\lg c$ as functions of $N_\mathrm{tracer}$.
  The diamonds show the total uncertainties calculated from the variances
  for similar distributions to those in the left panels. The thin gray curves show the 
  statistical uncertainties estimated using the stacked halos. The differences between the
  diamonds and the corresponding gray curves give the estimated systematic uncertainties,
  which are combined with the estimated statistical uncertainties to give the total
  uncertainties shown as the thick black curves. The filled (open) diamonds and
  solid (dashed) curves are for the default (flat) prior on $c$.
  \vspace{2mm}
  }
  \label{fig:esti_N}
\end{figure*}

The left panels of \reffig{fig:esti_N} show the statistics on the deviations
of the best-fit values, $\lg M_\mathrm{esti}$ and $\lg c_\mathrm{esti}$, from
the true values, $\lg M_\mathrm{true}$ and $\lg c_\mathrm{true}$, respectively, 
for the three sets of test halos when the default or flat prior on $c$ is used. 
It can be seen that our method provides asymptotically unbiased estimates, 
with better precision ($\sim 1/\sqrt{N_\mathrm{tracer}}$) for a larger 
$N_\mathrm{tracer}$. As quantitative measures of the total uncertainties, 
we use
\begin{align}
    \bar\sigma_{\mathrm{tot},\lg M}&\equiv 
    \sqrt{\mathrm{var}(\lg M_\mathrm{esti}-\lg M_\mathrm{true})}\,,\\
    \bar\sigma_{\mathrm{tot},\lg c}&\equiv
    \sqrt{\mathrm{var}(\lg c_\mathrm{esti}-\lg c_\mathrm{true})}\,,
\end{align}
where the variances are obtained from the corresponding distributions
similar to those shown in the left panels of \reffig{fig:esti_N}. 
The filled (open) diamonds in the right panels of this figure show the 
$\bar\sigma_{\mathrm{tot},\lg M}$ and $\bar\sigma_{\mathrm{tot},\lg c}$ for 
$N_\mathrm{tracer}=10$, 20, 40, 80, and 160 when the default (flat) prior on $c$ is 
used. Similar to the example in \refsec{sec:single}, using the default prior on $c$
can significantly improve the precision of the estimated $\lg c$ and mildly refine 
that of the estimated $\lg M$, especially when $N_\mathrm{tracer}$ is small.
Note that the estimated $\lg M$ is approximately twice more
precise than the estimated $\lg c$ for a flat prior on $c$. This result can be 
understood because in determining the depth of the potential as revealed by the tracers,
the halo mass plays the primary role, whereas the concentration is only a secondary 
factor. 

\subsection{Systematic uncertainties} \label{sec:systematics}

Whereas the basic assumptions of our method are reasonable and justified
in Appendix \ref{sec:equilibrium}, deviations from them are expected
when detailed aspects of the underlying cosmological simulation are
considered. For example, the density profile of a halo may not be of
the exact NFW type or not even strictly spherical. Some satellites
may have been accreted as a group and stay correlated in phase space, 
so they cannot be treated as independent tracers following a general
steady-state DF. Most importantly, variations in halo formation history 
may lead to different degrees of relaxation for satellite dynamics, 
and hence, varying deviations from a steady state. 
Such halo-to-halo scatter would violate strict scaling of internal 
dynamics for different halos. All deviations from our assumptions 
result in systematic uncertainties, which we estimate below by extending 
the analyses in \refsec{sec:tests}.

For each test halo, we construct a corresponding stacked halo whose satellites 
follow our simulation-based DF exactly, and therefore, present no systematic 
uncertainties for our method. Let $(\rs,\vs)$ be the characteristic scales
for the test halo and $(\rs',\vs')$ be those 
for another halo in the sample described in \refsec{sec:simulation}. We
scale the $(\bm{r}',\bm{v}')$ of every satellite for the latter halo
to $(\bm{r}=(\rs/\rs')\bm{r}',\bm{v}=(\vs/\vs')\bm{v}')$. This procedure
is repeated for all the halos other than the test halo in the above sample.
The satellites for the test halo and those for all the scaled halos are 
assigned to the stacked halo. By construction, those satellites with 
$30\leq r\leq 300\ \kpc$ for this stacked halo exactly follow the DF in
\refeqn{eqn:dfselect} that applies to a halo having the same $(M,c)$ as the 
test halo. Therefore, the estimated $(M,c)$ for the stacked halo from our
method only have statistical uncertainties. 

Using the stacked halos corresponding to all the test halos in the sample 
described in \refsec{sec:simulation}, we perform similar analyses to those 
for the test halos as presented in \refsec{sec:tests}. The statistical 
uncertainties in $\lg M$ and $\lg c$ obtained for the stacked halos as functions 
of $N_\mathrm{tracer}$ for the default (flat) prior on $c$ are shown as the 
thin gray solid (dashed) curves in the right panels of 
\reffig{fig:esti_N}. Note that because there are a large number of satellites
for any stacked halo, we can extend the analyses to $N_\mathrm{tracer}>10^3$.
Note also that the statistical uncertainties in $\lg M$ and $\lg c$ for the 
stacked halos become asymptotically independent of the prior on $c$ for 
$N_\mathrm{tracer}\gtrsim 100$ and 400, respectively.

For $N_\mathrm{tracer}=40$, the thin gray (solid or dashed) curves in the right panels 
of \reffig{fig:esti_N} give $\bar\sigma_\mathrm{stats}=0.059$ (0.068) and 0.086 (0.18)
for $\lg M$ and $\lg c$, respectively, when the default (flat) prior on $c$ is used.
These results are very close to those for the example application to a single halo 
presented in \refsec{sec:single}, which shows that the statistical uncertainties obtained
from the Bayesian formalism of our method are reliable.

Taking the statistical uncertainties $\bar\sigma_\mathrm{stats}$ in $\lg M$ and 
$\lg c$ for the test halos to be the same as those for the stacked halos shown as
the thin gray (solid or dashed) curves in the right panels of \reffig{fig:esti_N}, 
we can estimate the systematic uncertainties $\bar\sigma_\mathrm{sys}$ for the test 
halos by fitting 
\begin{equation}
  \bar\sigma^2_\mathrm{tot} = \bar\sigma^2_\mathrm{stats} + \bar\sigma^2_\mathrm{sys}
\end{equation}
to the total uncertainties $\bar\sigma_\mathrm{tot}$ 
shown as the (filled or open) diamonds in the right panels of the same figure for
$N_\mathrm{tracer}=10$, 20, 40, 80, and 160. We find that a good fit is obtained with
$\bar\sigma_\mathrm{sys}\approx 0.03$ (0.035) and 0.04 (0.07) dex for 
$\lg M$ and $\lg c$, respectively, when the default (flat) prior on $c$ is used.
Note that due to the lack of test halos with more than 160 satellites, we cannot
obtain reliable estimates of the total uncertainties for $N_\mathrm{tracer}>160$.
Using the above estimates for $\bar\sigma_\mathrm{stats}$ and $\bar\sigma_\mathrm{sys}$,
however, we can calculate $\bar\sigma_\mathrm{tot}$ for a wider range of 
$N_\mathrm{tracer}$ values. The results are shown as the thick black (solid or dashed) 
curves in the right 
panels of \reffig{fig:esti_N}. It can be seen that the statistical uncertainties
$\bar\sigma_\mathrm{stats}$ dominate for $N_\mathrm{tracer}<100$, whereas the
systematic uncertainties $\bar\sigma_\mathrm{sys}$ become non-negligible 
for $N_\mathrm{tracer}\sim 100$ and dominate for $N_\mathrm{tracer}\gtrsim 200$.
Therefore, the optimal number of tracers for our method is 
$N_\mathrm{tracer}\sim 100$. In general, the systematic uncertainties may be 
ignored for $N_\mathrm{tracer}< 100$, especially when the observational errors are 
taken into account. Nevertheless, it is always useful to point out the intrinsic 
limitation of a method by providing the systematic uncertainties.

Here the systematic uncertainties are estimated based on a sample of isolated halos. A question arises when the MW is considered because its massive neighbor
M31 is relatively close ($\sim 780$~kpc). However, \cite{Li2017} showed that the influence of a massive neighbor on satellite kinematics of a host halo is small when the neighbor is at a distance exceeding 3 times its virial radius (see their Figure~13). Based on this
criterion, M31 does not have a large effect on the kinematics of the MW satellites.
So we expect that our method can be applied to the MW with no large additional systematic uncertainties. Nevertheless, the systematic uncertainties of our method when applied to binary halos merit further detailed investigation.

\subsection{Dependence on kinematic data} \label{sec:info_used}

The above tests use the full set of kinematic data on the tracers as input.
In view of the difficulty in obtaining accurate measurement of radial and
tangential velocities, it is useful to study the dependence of the precision of
our method on individual kinematic observables, thereby providing guidance on 
effective inclusion of tracers with incomplete data.

\begin{figure}[htbp]
  \centering
  \includegraphics[width=0.45\textwidth]{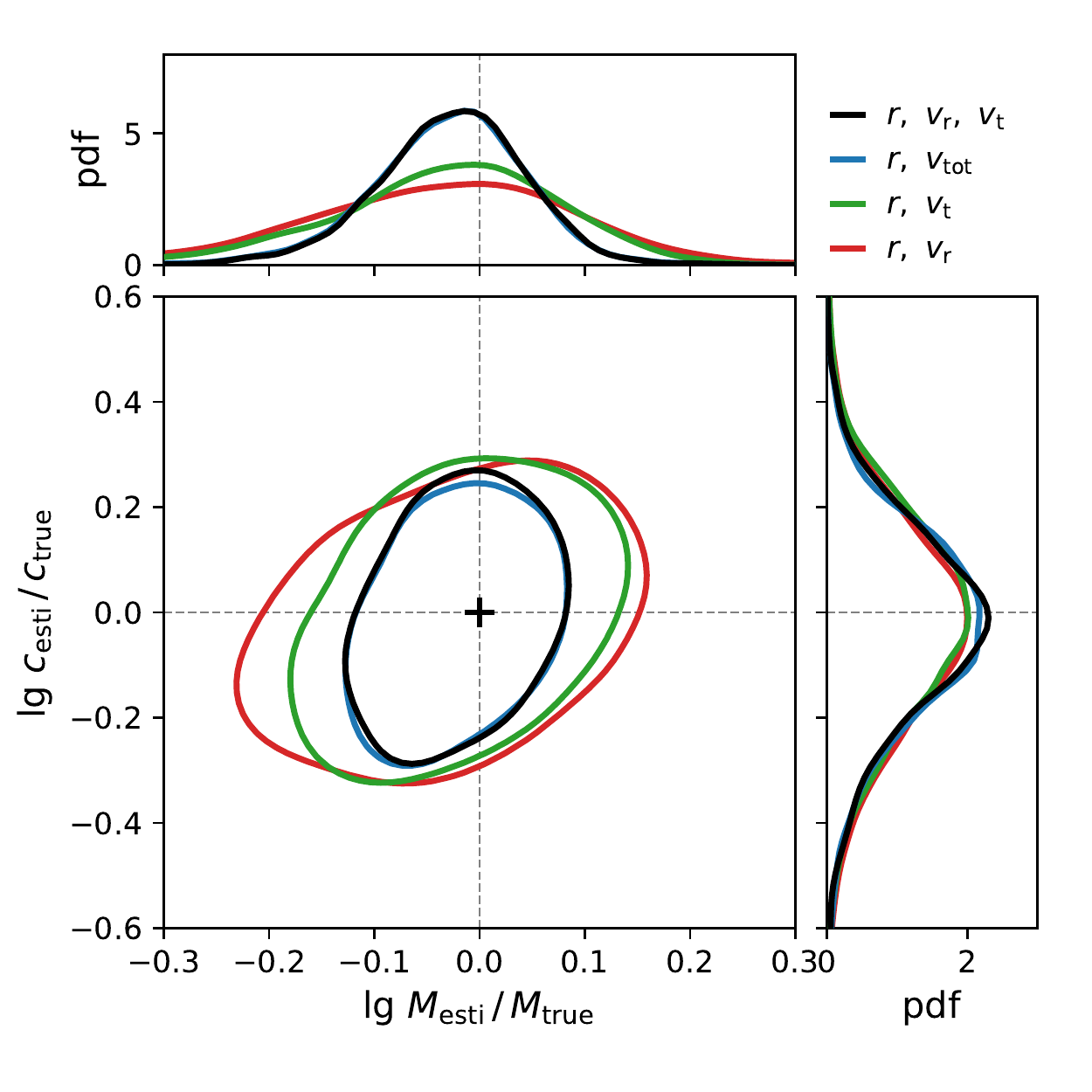}
  \caption{
  Tests of dependence of the precision on different sets of data for the halo sample.
  A total of 40 tracers per halo are used with a flat prior on $c$.
  The black, blue, green, and red curves show the $1\sigma$ confidence contours for the
  joint distribution of $\lg(M_\mathrm{esti}/M_\mathrm{true})$ and 
  $\lg(c_\mathrm{esti}/c_\mathrm{true})$ from the data on
  $(r,\vr\,,\vt)$, $(r,\vtot)$, $(r,\vt)$, and $(r,\vr)$, respectively.
  The same color coding is used for the marginalized distributions of
  $\lg(M_\mathrm{esti}/M_\mathrm{true})$ and 
  $\lg(c_\mathrm{esti}/c_\mathrm{true})$ shown in the top and right panels, respectively.
  Note that $\vtot$ cannot be observed directly, but is calculated from
  $\vr$ and $\vt$.
  }
  \label{fig:esti_info}
\end{figure}

Using 40 tracers per halo with a flat prior on $c$,
we apply our method to our test halo sample with data on
$(r,\vr\,,\vt)$, $(r,\vt)$, and $(r,\vr)$, respectively
(see \refsec{sec:incomp_info} for the appropriate DF in the latter two cases).
The resulting distributions of $\lg(M_\mathrm{esti}/M_\mathrm{true})$ and 
$\lg(c_\mathrm{esti}/c_\mathrm{true})$ are shown in \reffig{fig:esti_info}.
As expected, the data on $(r,\vr\,,\vt)$ provide significantly more precise 
estimate of $\lg M$ than the data on either $(r,\vt)$ or $(r,\vr)$.
In addition, because $\vt$ effectively represents two velocity components 
perpendicular to the radial direction, it provides tighter constraints on
$\lg M$ than $\vr$. Note also that the estimated $\lg c$ is insensitive to
which set of data among $(r,\vr\,,\vt)$, $(r,\vt)$, and $(r,\vr)$ is used.
We have also done tests with the data on $r$ only (not shown), and find that
spatial position provides very weak constraints compared with velocity. Therefore,
with position only, a very large sample of tracers is required to provide meaningful
estimates of halo properties.

As mentioned in \refsec{sec:incomp_info}, previous studies 
\citep[e.g.,][]{Binney2008,Wojtak2008} showed that the mass distribution of a system 
is mainly determined by the energy distribution of its constituents but
is insensitive to their velocity anisotropy or angular momentum distribution.
This result is supported by the work of \citet{Callingham2018}, who found that
using the probability density function in either the $E$ or $(E,L)$ space constrains 
the halo mass equally well. To further test the above result, we combine
$\vr$ and $\vt$ into $\vtot$ and apply our method using the data on $(r,\vtot)$ 
(see \refsec{sec:incomp_info} for the appropriate DF). The corresponding distributions 
of $\lg(M_\mathrm{esti}/M_\mathrm{true})$ and 
$\lg(c_\mathrm{esti}/c_\mathrm{true})$ are shown in \reffig{fig:esti_info}. It can be seen
that the data on $(r,\vtot)$ without distinguishing $\vr$ and $\vt$ provide essentially 
the same constraints on halo properties as the data on $(r,\vr\,,\vt)$.
Therefore, whereas halo-to-halo scatter gives rise to systematic uncertainties of our
method, the scatter in the velocity anisotropy contributes very little, which is
consistent with the very small systematic uncertainties of our method.
We note however, that $\vtot$ cannot be measured directly. So although the influence of
$\vr$ and $\vt$ enters our method mostly in the form of $\vtot$, this insensitivity 
to the velocity anisotropy only holds when both $\vr$ and $\vt$ are available.
Otherwise, estimates of the halo mass are inevitably affected by the 
uncertainty in the velocity anisotropy (e.g., \citealt{Wolf2010a,Watkins2010a}).

\subsection{Comparison with other methods} \label{sec:comparison}

As mentioned in \refsec{sec:intro}, many methods have been proposed to estimate the halo 
mass based on dynamical modeling, especially for the MW. Here we compare our method with
the oPDF method\footnote{
  The Python implementation of the oPDF method is publicly available at 
  \url{https://github.com/Kambrian/oPDF}.} \citep{Han2016b}.
The oPDF method makes the minimum assumption of steady-state tracer dynamics in
a static potential and is representative of a large category of steady-state methods, including those based on the Jeans equation and those based on the Jeans theorem such as Schwarzchild modeling.
In turn, the Jeans equation method has been tested against many other methods
and show close (or slightly better) precision compared to the
caustic or virial method (e.g., \citealt{Rines2006,Old2015,Armitage2019}).

Under the steady-state assumption, the oPDF method redistributes the radial position of each tracer 
along its orbit for a trial static potential according to $p(r|E,L)$ given by \refeqn{eqn:prel}. 
The difference between the resulting radial distribution of the tracer 
sample and the observed distribution is minimized using a likelihood function to infer
the correct potential. In principle, the oPDF method can be applied to potentials of 
arbitrary (including non-spherical) form. In the comparison below, we take the
potential of a test halo as given by its fitted NFW profile and use the radial
likelihood estimator of the oPDF method (with 10 radial bins, see \citealt{Han2016b} 
for details) to infer the corresponding halo mass and concentration.

In addition, to further illustrate the difference between our method, which is based on 
the DF, $f(E,L)$, 
in the phase space of $(\bm{r},\bm{v})$, and the method of \citet{Li2017} (see also
\citealt{Callingham2018}), which is based on the probability density function, $p(E,L)$,
in the $(E,L)$ space, we compare these two methods using the self-consistent $f(E,L)$ 
and $p(E, L)$ of our method. Note that our $p(E,L)$ can be used to infer both the halo
mass and concentration, whereas the method of \citet{Li2017} was developed for estimating
the halo mass only. To mimic that method more closely, we also apply it assuming the 
average $\lg M$--$\lg c$ relation for our halo sample (see Equation \ref{eqn:lgc}) so that only 
the halo mass needs to be inferred.

\begin{figure}[htbp]
  \centering
  \includegraphics[width=0.45\textwidth]{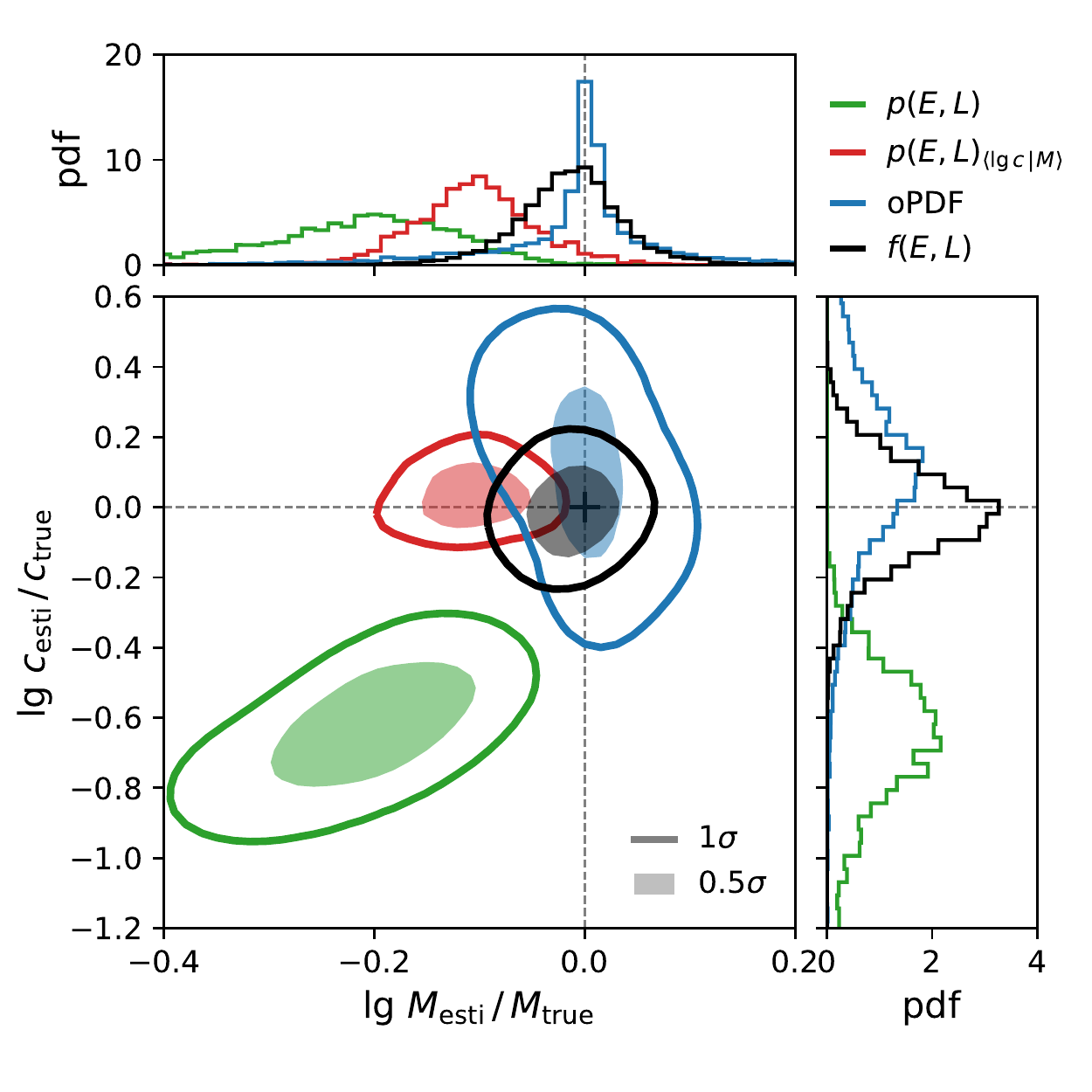}
  \caption{
  Comparison of different methods applied to the halo sample with 80 tracers per halo.
  A flat prior on $c$ is used except for the method based on $p(E,L)$ with the average
  $\lg M$--$\lg c$ relation (labeled $p(E,L)_{\langle\lg c|M\rangle}$). 
  The black, blue, green, and red curves (shaded areas) show the $1\sigma$ ($0.5\sigma$)
  confidence contours (regions) for the joint distribution of $\lg(M_\mathrm{esti}/M_\mathrm{true})$ 
  and $\lg(c_\mathrm{esti}/c_\mathrm{true})$ for our method (labeled $f(E,L)$), the oPDF
  method, the method based on $p(E,L)$, and the method based on 
  $p(E,L)_{\langle\lg c|M\rangle}$, respectively.
  The same color coding is used for the marginalized distributions of
  $\lg(M_\mathrm{esti}/M_\mathrm{true})$ and 
  $\lg(c_\mathrm{esti}/c_\mathrm{true})$ shown in the top and right panels, respectively.
  }
  \label{fig:esti_compare}
\end{figure}

Using 80 tracers per halo with a flat prior on $c$ (except for the $p(E,L)$ method
with the average $\lg M$--$\lg c$ relation), we apply the above methods to our halo sample 
and show their distributions of $\lg(M_\mathrm{esti}/M_\mathrm{true})$ and 
$\lg(c_\mathrm{esti}/c_\mathrm{true})$ in
\reffig{fig:esti_compare} and give the corresponding average values,
$\langle\lg(M_\mathrm{esti}/M_\mathrm{true})\rangle$ and 
$\langle\lg(c_\mathrm{esti}/c_\mathrm{true})\rangle$, as well as
the total uncertainties, $\bar\sigma_{\lg M}$ and $\bar\sigma_{\lg c}$,  
in \reftab{tab:method_compare}. It can be seen that both the oPDF method and our method 
are able to provide unbiased estimates of the halo mass and concentration. Note that
although the distribution of $\lg(M_\mathrm{esti}/M_\mathrm{true})$ for the oPDF method 
sharply peaks at zero, its long tails on both sides of the peak result in
a larger total uncertainty than that for our method. The oPDF method is also less precise
in estimating $\lg c$ than our method, although the precision can be improved in both
cases when the default prior on $c$ is used (not shown). A similar comparison with similar 
results to the above was shown in Figure 3 of \citet{Han2016b} for an idealized case.

\begin{table}[htbp]
\centering
\caption{
  Comparison of the methods as shown in \reffig{fig:esti_compare}.
}
\label{tab:method_compare}
\begin{tabular}{ccccc}
\toprule
 & $\big\langle \lg \frac{M_\mathrm{esti}}{M_\mathrm{true}} \big\rangle$ & $\sigma_{\lg M}$ 
 & $\big\langle \lg \frac{c_\mathrm{esti}}{c_\mathrm{true}} \big\rangle$ & $\sigma_{\lg c}$ \\
\midrule
                               $p(E, L)$ & $   -0.24$ & $    0.13$ & $   -0.68$ & $    0.23$\\
    $p(E, L)_{\langle\lg c|M\rangle}$    & $   -0.10$ & $    0.06$ & $       -$ & $       -$\\
                                    oPDF & $   -0.00$ & $    0.11$ & $   \ \ 0.14$ & $    0.41$\\
                               $f(E, L)$ & $   \bm{-0.01}$ & $ \bm{0.05}$ & $   \bm{-0.02}$ & $ \bm{0.14}$\\
\bottomrule
\end{tabular}
\end{table}

As already noted by \citet{Li2017}, the estimated halo mass based on $p(E,L)$ 
has an intrinsic bias. We have mentioned 
in \refsec{sec:intro} that this bias arises because $E$ is not directly observed.
For estimating halo properties based on the $(\bm{r},\bm{v})$ of tracers, the 
proper DF to use is $f(E,L)$. With $p(E,L)=8\pi^2LT_\mathrm{r}(E,L)f(E,L)$, 
a potential with a longer radial orbital period $T_\mathrm{r}(E,L)$ for the same 
$(\bm{r},\bm{v})$ is favored by $p(E,L)$. Because a shallower potential corresponding
to a lower halo mass allows a tracer to have a more distant apocenter, and hence, 
a longer $T_\mathrm{r}(E,L)$, the method based on $p(E,L)$ underestimates
the halo mass. \reffig{fig:esti_compare} and \reftab{tab:method_compare} indeed 
show such a bias. Note that both the bias and the total uncertainty for the estimated
halo mass are reduced significantly when the average $\lg M$--$\lg c$ relation is
used in the method based on $p(E,L)$ to mimic the original method of \citet{Li2017}.
The typical bias of $\sim -0.1$ dex in this case is also consistent with the results of
\citet{Li2017}.

Although the method based on $p(E,L)$ gives biased estimates of the halo mass, it
qualitatively illustrates the constraining power of the orbital distribution in the
$(E,L)$ space, whereas the oPDF method shows that of the radial distribution $p(r|E,L)$ along
each orbit. As indicated by the nearly orthogonal $1\sigma$ confidence contours of 
these two methods in \reffig{fig:esti_compare}, the tightest constraints on halo
properties are provided by combining both the orbital distribution and the radial
distribution along each orbit, which is exactly done by the DF, $f(E,L)$, of our
method (see Equation \ref{eqn:frvprel2}). Therefore, our method is more precise than
the oPDF method as shown above, and perhaps, has the best performance among
the currently existing methods.

\section{Applications to the MW and beyond} \label{sec:discussion}

A key motivation for this work is to estimate the mass and its detailed 
distribution for the MW halo. These properties are crucial to many
astrophysical studies, but remain rather uncertain
(see Wang et al.\ 2019 in preparation for a comprehensive summary of recent measurements).
Whereas satellite galaxies are the best tracers for the mass distribution of 
the outer MW halo and the total halo mass \citep{Han2019}, their use for mass estimates
has been limited by their small sample size and poor kinematic data until recently.
Deep sky surveys have doubled the number of known MW satellites over the past 
several years (see \citealt{Simon2019} for a review). In addition,
the unprecedented precision of Gaia \citep{GaiaCollaboration2018a} has enabled
better proper motion measurement for the MW satellites
(e.g., \citealt{GaiaCollaboration2018,Kallivayalil2018,Pace2018b,Fritz2018a}).
With the above improvements of the data, we apply our method to the MW.
The details are presented in a separate paper (Li et al.\ 2019, in preparation),
and the main results are summarized below.

Using 28 satellites with Gaia DR2 proper motion data and the DF model based on halos in the Eagle simulation, we obtain
$M=1.23_{-0.18}^{+0.21}\times 10^{12}\msun$ and $c=9.4_{ -2.1}^{ +2.8}$ for
the MW halo. Both the selection function and measurement errors are taken
into account and treated rigorously within the Bayesian statistical framework
of our method. With a $\sim 20\%$ uncertainty, our estimated MW halo mass is 
currently the most precise. 
The systematic error due to halo-to-halo scatter is small compared to the the current
statistical uncertainty.
The inferred halo mass is consistent with 
recent measurements using satellites (\citealt{Li2017,Callingham2018,Patel2018}),
stars (e.g., \citealt{Zhai2018,Deason2019}),
and globular clusters (e.g., \citealt{Sohn2018,Watkins2018,Vasiliev2018}).
This mass estimate can be further improved if multiple tracer populations (e.g., satellites and halo stars) are used in combination
(Li et al.\ 2019, in preparation).
In addition, our DF model along with the
estimated halo mass can be used to constrain the kinematics of distant satellites, 
so that their orbits, and hence the assembly history of the MW, may be better 
inferred.

In principle, our method can be applied to halos of any mass or concentration,
including galaxy groups or clusters, so long as these systems follow the same DF 
when scaled by the corresponding $\rs$ and $\vs$. 
Galaxy clusters, however, are further from 
equilibrium compared to galactic halos. The larger halo-to-halo scatter for the 
former is expected to result in a larger intrinsic uncertainty. In addition,
substructures in more massive halos tend to move along more radial orbits. This
feature may be problematic when there are no data on the associated proper motion.
As discussed in \refsec{sec:info_used}, although our method is insensitive to the
velocity anisotropy when the full kinematic information of the tracers is available,
lack of either $\vr$ or $\vt$ is expected to weaken this insensitivity, thereby
increasing the intrinsic uncertainty due to the halo-to-halo scatter from the velocity 
anisotropy. In practice, the above issues can be mitigated by constructing the
appropriate DF from template halos in the relevant mass range.

For application to distant galaxies, we must deal with more complex observations
\citep[e.g.,][]{Biviano2006,Wojtak2018,Lange2019,Pratt2019}. For example, when only the projected position 
and line-of-sight velocity are observed for satellites, we need to marginalize the DF 
to model the projected phase space \citep[see e.g.,][]{Dejonghe1992}. In addition, there are other problems common to
nearly all methods, such as foreground and background contamination (interlopers),
incompleteness (e.g., caused by fiber collision),
mis-centering, modeling of the infall region near the halo boundary,
and influence of the large-scale structure. All of the above issues require careful 
treatment and merit future studies.

\section{Discussion and Conclusions} \label{sec:conclusion}

We have proposed a new method to estimate the properties, especially the mass,
of a halo from the phase-space distribution of its satellites. The DF in
phase space is constructed directly from a cosmological simulation assuming 
similarity of internal dynamics for different halos. Within the fully Bayesian 
framework of our method, which is unbiased and efficient, we are able to infer 
both the halo mass and the concentration, and treat various observational effects, 
including the selection function, incomplete data (e.g., lack of proper motion),
and observational errors (see Li et al.\ 2019, in preparation) 
in a rigorous manner.

We have tested the validity and accuracy of our method with mock samples.
Making full use of the DF in phase space, our method achieves better precision
than the oPDF method, which is representative of 
a large family of pure steady-state methods, 
including those based on the Jeans equation and Schwarzschild modeling.
Because our new method makes use of the mass-dependent distribution of orbits in addition to the steady-state distribution along each orbit, 
we are able to slightly reduce the stochastic systematic uncertainty~\citep{Wang2017b, Wang2018} that represents the information limit of pure steady-state methods.
In the ideal case without observational errors, we are able to constrain the halo mass 
at the 20\% level ($\sim 0.08$ dex in $\lg M$) with only 20 satellites. The systematic 
uncertainty is $\sim 8\%$ ($\sim 0.035$ dex in $\lg M$) for the halo mass and
$\sim 16\%$ ($\sim 0.07$ dex in $\lg c$) for the concentration with a flat prior on
the latter. 

These results are comparable to the $\sim 25\%$ and $40\%$ systematic uncertainties\footnote{There is a subtle difference between the systematic uncertainties quoted in \citet{Wang2017b,Wang2018} and ours. 
The former uncertainty was based on the extent of the $1\sigma$ confidence region in the two-dimensional $(M,c)$ space, 
while we adopt the marginalized one-dimensional uncertainty 
which is expected to be about half of the corresponding two-dimensional extent.} found in \citet{Wang2017b, Wang2018} for the halo mass and concentration, respectively, when dark matter particles from simulations were used as tracers. 
On the other hand, the systematic uncertainties when using star particles as tracers can be as high as $\sim 300\%$ according to \citet{Wang2017b, Wang2018}, much larger than what we found using satellite tracers. 
These results can be understood as satellite galaxies are nearly unbiased phase-space tracers of dark matter particles, 
while halo stars remain highly phase-correlated after getting stripped from their progenitors. 
We leave more detailed comparisons and discussions of the different tracers to a separate paper \citep{Han2019}.

A major application of our method is to estimate the MW halo mass.
Using the kinematic data of satellites updated by Gaia,
we obtain a mass of $1.23\times 10^{12}\msun$ with a $\sim 20\%$ uncertainty
for the MW halo, which is consistent with other recent estimates from various tracers
(e.g., \citealt{Callingham2018,Patel2018,Deason2019,Zhai2018,Sohn2018,Watkins2018,Vasiliev2018}).
The significantly lower mass $M=0.71\times 10^{12}\msun$ obtained by \citet{Eadie2018}
likely reflects the specific assumptions in the underlying DF model and gravitational potential.
A detailed report is given elsewhere (Li et al.\ 2019, in preparation).
Our method can also be applied to other halos including galaxy groups or clusters.
We plan to carry out such followup studies in the future.

We have used a mock galaxy sample, which was generated from the SAM based on
a cosmological simulation, to construct the DF and validate our method.
Whereas the validity of our method has been demonstrated for this SAM sample,
whether this sample represents actual galaxies is a concern when we apply our method
to the MW and other real systems. Simulations and models of galaxy formation are
plagued by the poorly-understood processes
in baryon-dominated regions. In general, the presence of a stellar disc and 
adiabatic contraction of a halo would alter the potential of the inner halo
and enhance the tidal disruption of substructures. Simulations, however,
show that satellites in the outer halo are less affected
(e.g., \citealt{Zhu2016a,Sawala2017,Richings2018}). In particular,
satellite kinematics is largely unchanged by baryonic physics
for radii exceeding one quarter of the virial radius \citep{Sawala2017,Richings2018}. 
In addition, \citet{Gifford2013} reported that halo mass estimates using satellites 
are insensitive to variation of the dynamical friction applied to the ``orphan'' 
galaxies in the SAM. Therefore, the SAM galaxy sample provides a reasonable means
to represent the satellite kinematics in the outer halos of real galaxies,
and our method can be used to estimate the properties of these systems from the
actual satellite data. Nevertheless, the appropriateness of the SAM galaxy sample 
warrants more in-depth and systematic studies.
More detailed comparison of halo mass estimates based on SAM and hydrodynamic simulations 
will be given in Li et al.\ 2019, in preparation.

Finally, the validity of our method supports the similarity of internal dynamics
of different halos, at least for the SAM galaxy sample. The corresponding DF
can provide some insights into and facilitate analytical studies of the dynamical 
state of halos.

\acknowledgments

We thank Rados\l{}aw Wojtak for communications on DF models, and Carlos S. Frenk, 
Lu Li, Ting Li, Zhengyi Shao, and Jiajun Zhang for helpful discussions.
This work was supported in part by the National Natural Science Foundation of China
[11533006, 11621303, 11890691, 11655002, 11873038],
National Key Basic Research and Development Program of China (No.\ 2018YFA0404504),
the US Department of Energy [DE-FG02-87ER40328 (UM)],
the National Program on Key Basic Research Project (Grant No.\ 2015CB857003), 
the Science and Technology Commission of Shanghai Municipality [16DZ2260200],
and JSPS Grant-in-Aid for Scientific Research JP17K14271.

This work made use of the High Performance Computing Resource 
in the Core Facility for Advanced Research Computing at Shanghai Astronomical Observatory,
and the computing facilities at Department of Astronomy, School of Physics and Astronomy, 
Shanghai Jiao Tong University.

\textit{Software:} 
  Numpy \citep{Walt2011}, 
  Scipy \citep{Oliphant2007},
  Matplotlib \citep{Hunter2007},
  Astropy \citep{AstropyCollaboration2013},
  oPDF \citep{Han2016b}

\appendix \label{appendix}

\section{Relation between phase space and $(E,L)$ space}
\label{sec:felpel}

Here we derive the relation between the DF $f(E,L)$ in the phase space of 
$(\bm{r},\bm{v})$ and the probability density function $p(E,L)$ in the $(E,L)$
space. From the assumed spherical symmetry, we can write
\begin{equation}
f({\bm{r}},{\bm{v}})d^3{\bm{r}}d^3{\bm{v}}=f(r,\vr\,,\vt)8\pi^2r^2\vt\,drd\vr\,d\vt.
\end{equation}
Because we can fully specify a tracer by the $(E,L)$ of its orbit and its radius,
$r$, we have
\begin{equation}
f(r,\vr\,,\vt)8\pi^2r^2\vt\,drd\vr\,d\vt=p(r|E,L)p(E,L)drdEdL,
\label{eqn:frvprel}
\end{equation}
where $p(r|E,L)$ is the probability density function of $r$ for a specific
set of $(E,L)$. Using the Jacobian for the transformation
of variables from $(r,E,L)$ to $(r,\vr\,,\vt)$, 
\begin{equation}
\frac{\partial(r,E,L)}{\partial(r,\vr\,,\vt)}=r|\vr|\,,
\end{equation}
we rewrite \refeqn{eqn:frvprel} as
\begin{equation}
f(r,\vr\,,\vt)=\frac{|\vr|}{8\pi^2L}p(r|E,L)p(E,L),
\label{eqn:frvprel3}
\end{equation}
which is \refeqn{eqn:frvprel2} in \refsec{sec:method}.

It is useful to define a radial phase angle
\begin{equation}
\theta\equiv\frac{2\pi}{T_{\mathrm{r}}(E,L)}\int_{\rperi}^r\frac{dr'}{\vr(r',E,L)},
\label{eqn:theta}
\end{equation}
where
\begin{equation}
\vr(r',E,L)=\pm\sqrt{2[E-\Phi(r')]-L^2/r'^2}\,,
\end{equation}
and the plus (minus) sign is for motion away from (towards) the pericenter.
\refeqn{eqn:prel} can now be rewritten as
\begin{equation}
p(r|E,L)dr=p(\theta|E,L)d\theta=\frac{d\theta}{2\pi}.
\label{eqn:ptheta}
\end{equation}
Thus, $\theta$ is uniformly distributed over $[-\pi,\pi]$ under our steady-state 
assumption.

Note that when a sample of satellites with a limited radial range of
$r_{\mathrm{min}}\leq r\leq r_{\mathrm{max}}$ is used to construct $f(E,L)$,
the lower and upper limits for the radial integral in \refeqn{eqn:tr} for
$T_{\mathrm{r}}(E,L)$ should be replaced by $\mathrm{max}\{\rperi,r_{\mathrm{min}}\}$
and $\mathrm{min}\{\rapo,r_{\mathrm{max}}\}$, respectively \citep{Wojtak2008}.
Similar adjustments should be made for $\theta$ as well.
Nevertheless, it is important to recognize that the DF $f(E,L)$
is not subject to any radial limit.
As the radial range changes, both $p(E,L)$ and $p(r|E,L)$ vary in a complementary way 
to keep their product invariant, so long as the entire system is in a steady state.

\section{Validity of assumptions} \label{sec:equilibrium}

Our assumed NFW profile for halos is a well-known result 
in the framework of hierarchical structure formation \citep{Navarro1996}.
This profile is a good fit for our halo sample, with 
differences of $\lesssim 3\%$ between the fitted halo masses and 
the true values.

As a check on the assumed steady state of satellite kinematics,
we show that satellites in our template sample are uniformly distributed in phase angles along
the orbits (see e.g., \citealt{Binney2008,Han2016b}). In other words, 
the probability density function
$p(\theta|E,L)$ is $1/(2\pi)$ as in \refeqn{eqn:ptheta}.
Note that satellites in our template sample have $r<\rlim=25\rs$. So
the upper limit for the radial integral in \refeqn{eqn:tr} for $T_{\mathrm{r}}(E,L)$,
as well as the maximum $r$ for the definition of $\theta$ in \refeqn{eqn:theta},
should be the smaller of $\rlim$ and $\rapo$.

\begin{figure*}[htbp]
  \centering
  \includegraphics[width=0.45\textwidth]{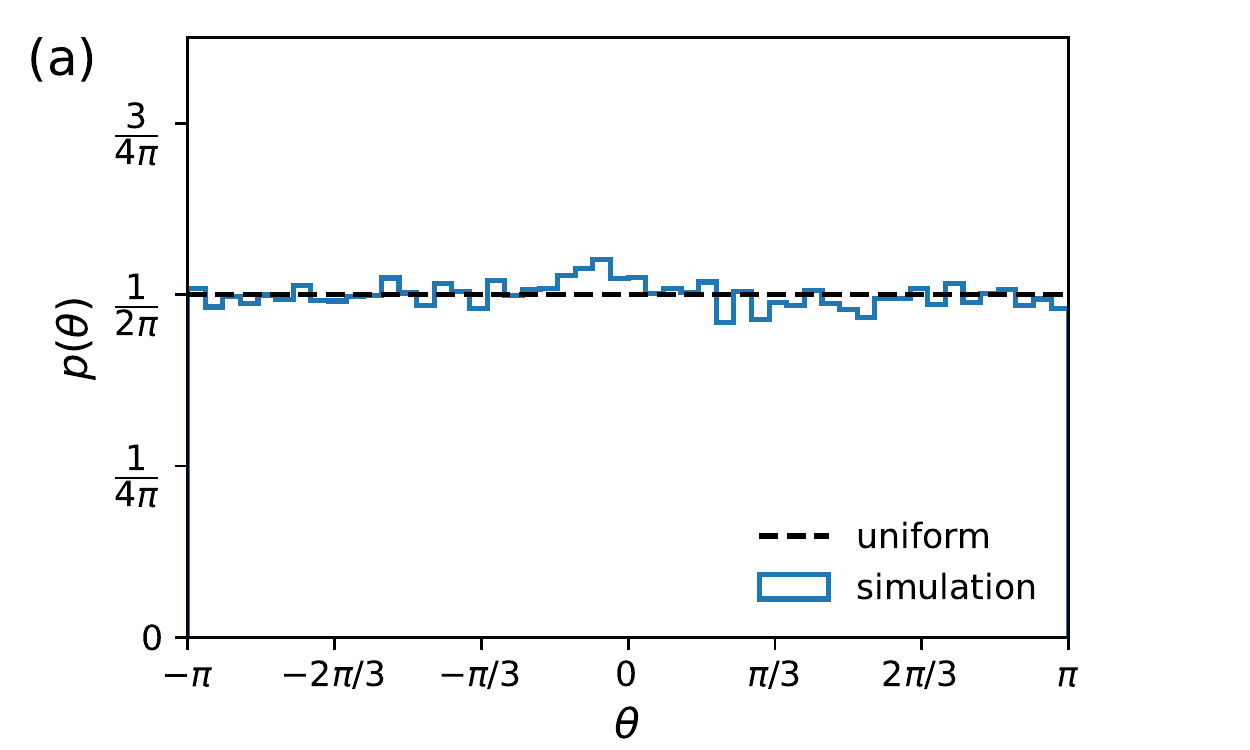}
  \includegraphics[width=0.45\textwidth]{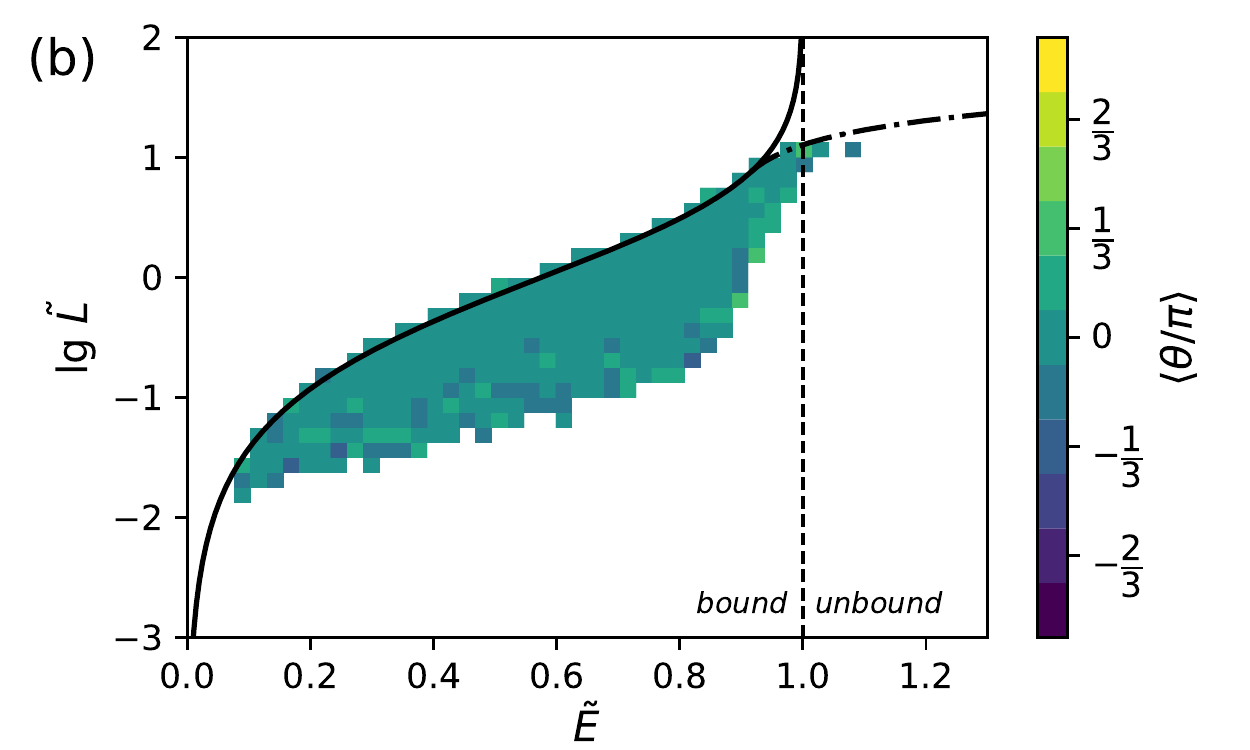}
  \includegraphics[width=0.45\textwidth]{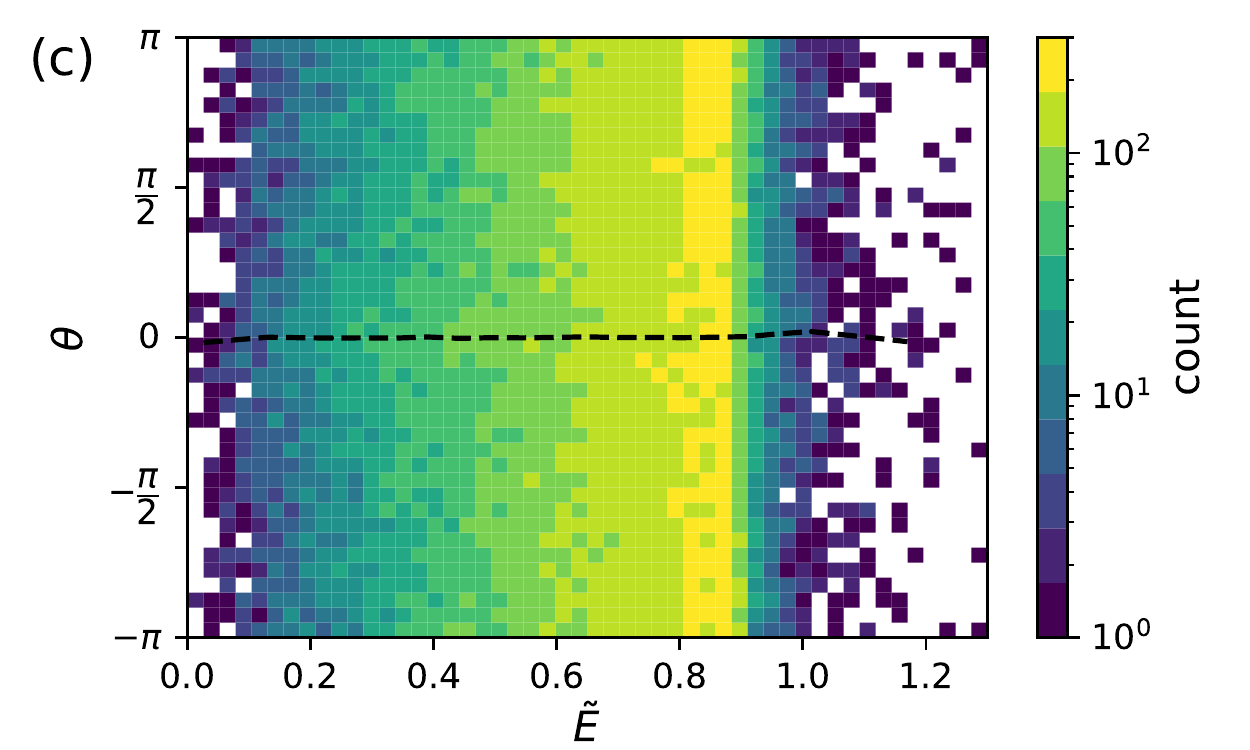}
  \includegraphics[width=0.45\textwidth]{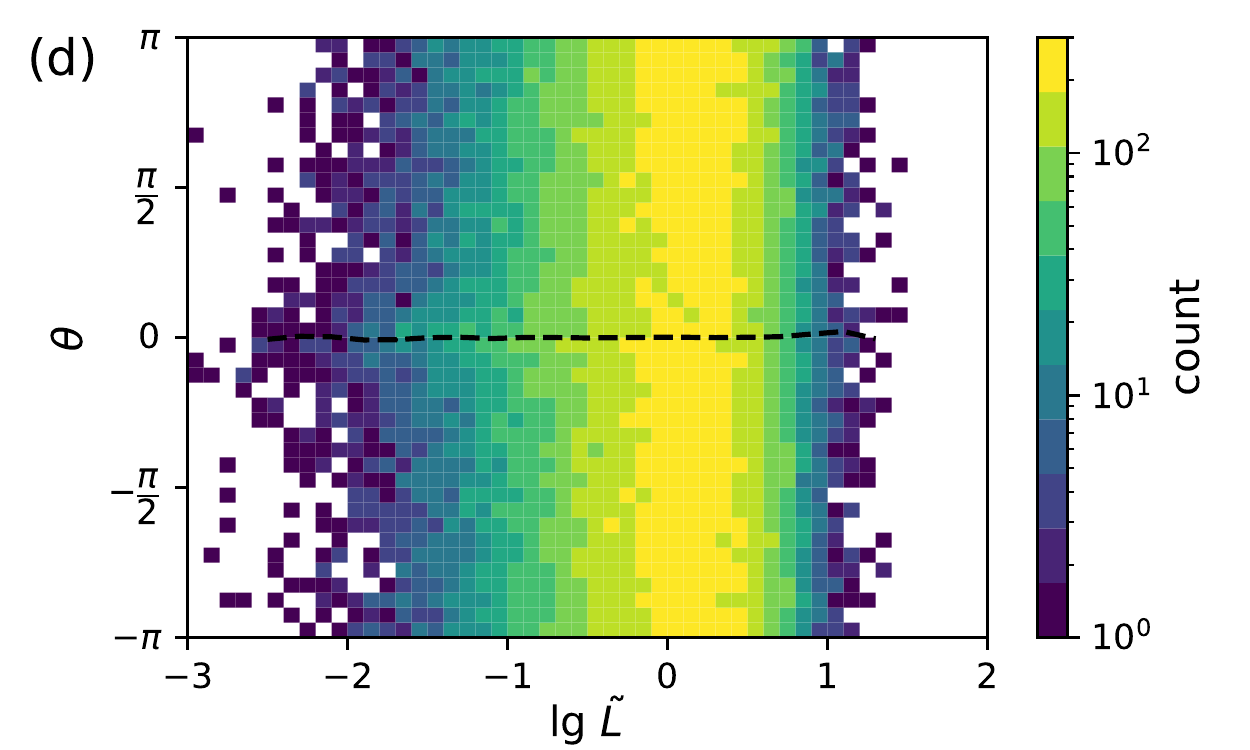}
  \caption{
    Distributions of radial phase angle $\theta$ for satellites in the template sample.
    (a) The distribution $\tilde p(\theta)=\int\tilde p(\theta|\tilde E,\tilde L)
    \tilde p(\tilde E,\tilde L)d\tilde Ed\tilde L$.
    (b) The average value 
    $\langle\theta(\tilde E,\tilde L)\rangle=\int\theta\tilde p(\theta|\tilde E,\tilde L)d\theta$
    as a function of $(\tilde E,\tilde L)=(E/\vs^2,L/(r\vs))$. Only pixels with more than 20 
    satellites are displayed. The solid (dot-dashed) curve shows the maximum $\tilde L$ 
    as a function of $\tilde E$ for any radius (for $r<25\rs$). (c) Numbers of satellites in 
    various $(\theta,\tilde E)$ bins. (d) Numbers of satellites in various $(\theta,\tilde L)$ bins.
    The dashed curve in (c) or (d) shows the average $\theta$ for bins with a fixed $\tilde E$ or 
    $\tilde L$, respectively.
    \vspace{2mm}
  }
  \label{fig:phase_distr}
\end{figure*}

It is difficult to show $p(\theta|E,L)$ directly, which requires 4D display. 
Instead, we show in \reffig{fig:phase_distr} the distribution 
$\tilde p(\theta)=\int\tilde p(\theta|\tilde E,\tilde L)
\tilde p(\tilde E,\tilde L)d\tilde Ed\tilde L$, and
the average value $\langle\theta(\tilde E,\tilde L)\rangle=
\int\theta\tilde p(\theta|\tilde E,\tilde L)d\theta$
as a function of $(\tilde E,\tilde L)$, for the satellites in our template sample.
It can be seen that $\tilde p(\theta)\approx 1/(2\pi)$ and 
$\langle\theta(\tilde E,\tilde L)\rangle\approx 0$ for all sets of 
$(\tilde E,\tilde L)$, as expected from $\tilde p(\theta|\tilde E,\tilde L)=1/(2\pi)$.
In addition, we show in \reffig{fig:phase_distr} numbers of satellites in 
various $(\tilde\theta,\tilde E)$ or $(\tilde\theta,\tilde L)$ bins, which are also
in excellent agreement with a uniform distribution of $\theta$.

\begin{figure*}[htbp]
  \centering
  \includegraphics[width=0.9\textwidth]{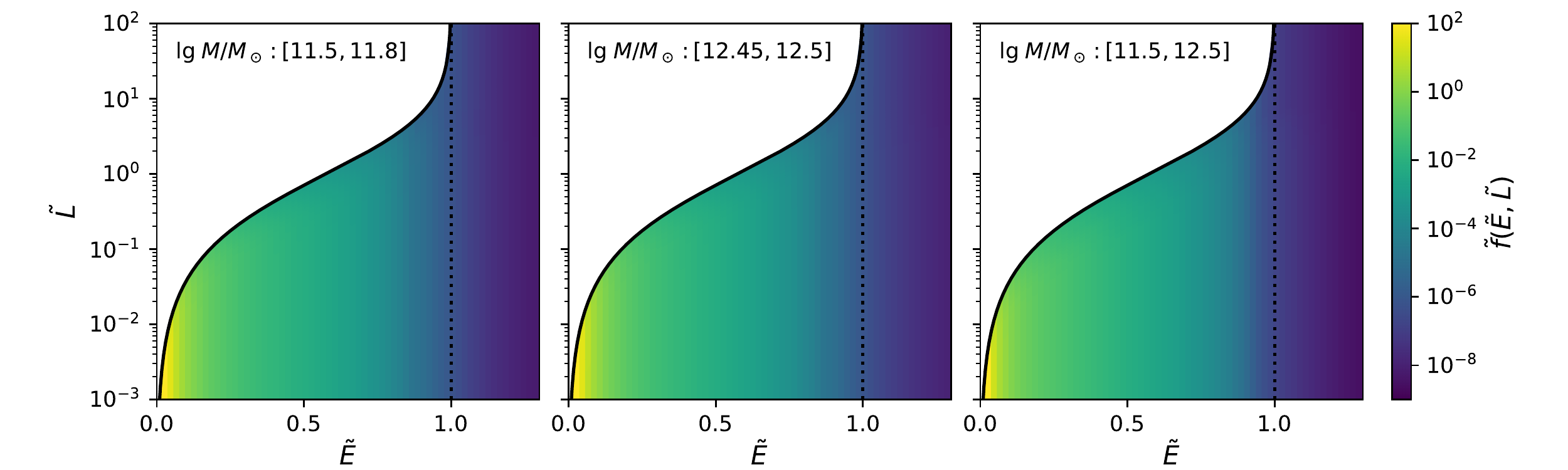}
  \caption{
Phase-space distribution functions $\tilde f(\tilde E,\tilde L)$ constructed from the template 
sample of satellites and its subsamples for halos in the indicated mass ranges. The left or middle 
panel contains $\approx  10^4$ satellites. The right panel is for the template sample of 104,315 
satellites and is the same as the DF shown in \reffig{fig:DF}.}
  \label{fig:mass_dep}
\end{figure*}

Finally, our assumption regarding the similarity of internal dynamics for different halos
is supported by the results of \citet{Li2017} and \citet{Callingham2018} on the probability 
density function $\tilde p(\tilde E,\tilde L)$ in the $(\tilde E,\tilde L)$ space. 
In addition, we show in \reffig{fig:mass_dep} the distributions 
$\tilde f(\tilde E,\tilde L)$ in the phase space 
of $(\tilde{\bm{r}},\tilde{\bm{v}})$ that are constructed from our template sample of satellites
and its subsamples for halos with $\lg M/M_\odot\in [11.5,12.5]$,
$[11.5,11.8]$, and $[12.45,12.5]$, respectively. 
It can be seen that all these distributions are nearly identical despite the large
differences in the halo mass range used.

\section{Details of constructing the DF} \label{sec:smooth}

We use the satellites in the template sample to construct a smooth DF in the
phase space of $(\tilde{\bm{r}},\tilde{\bm{v}})$. All the satellites have
$\tilde r\leq \tilde r_{\lim}=25$ relative to the centers of their host halos.
Because their $(\tilde E,\tilde L)$ are unevenly distributed in a sharp triangular
region (see e.g., \citealt{Li2017}), a fixed smoothing kernel in the $(\tilde E,\tilde L)$ 
space is not adequate. Instead, an adaptive smoothing procedure is needed.
We perform this procedure in the parameter space of $(\tilde E, j^2)$, where 
$j =\tilde L/\tilde L_{\max} (\tilde E) \in[0, 1]$ and $\tilde L_{\max} (\tilde E)$ is 
the maximum angular momentum for a satellite of energy $\tilde E$ in the template sample.
If the radius of the circular orbit for the energy $\tilde E$ is 
$\tilde r_{{\mathrm{cir}}} (\tilde E) \leq \tilde r_{\lim}$, then
$\tilde L_{\max} (\tilde E)$ is the angular momentum of the circular orbit,
$\tilde L_{\mathrm{cir}}(\tilde E)$,
and $j$ is the so-called orbital circularity. Clearly, this definition of
$\tilde L_{\max} (\tilde E)$ does not apply if 
$\tilde r_{{\mathrm{cir}}} > \tilde r_{\lim}$ or if the orbit is unbound with
$\tilde E > \tilde\Phi(\infty)=1$. For these two cases, we define 
$\tilde L_{\max} (\tilde E)$ as the product of $\tilde r_{\lim}$ and the total velocity
$\tilde v_{\mathrm{tot}}$ at this radius for the energy $\tilde E$, so that
$j\in[0,1]$ is again valid. Specifically, we have
\begin{equation}
  \tilde L_{\max} (\tilde E) = \begin{cases}
    \tilde L_{\mathrm{cir}}(\tilde E)\,,
       &\mathrm{for}\ \tilde r_{{\mathrm{cir}}} (\tilde E) \leq \tilde r_{\lim}\,,\\
    \tilde r_{\lim} \sqrt{2 [\tilde E - \tilde\Phi (\tilde r_{\lim})]}\,,
       &\mathrm{otherwise},
  \end{cases}
\end{equation}
which is shown as the dot-dashed curve in \reffig{fig:phase_distr}b.

The DF in \refeqn{eqn:fel} can be rewritten as
\begin{equation}
 \tilde f (\tilde E,\tilde L) = \frac{1}{4 \pi^2\tilde T_{\mathrm{r}}(\tilde E,\tilde L)
 \tilde L^2_{\max} (\tilde E)} \times \frac{d^2 N}{d\tilde E d j^2}.
\end{equation}
To obtain a smooth $\tilde f(\tilde E,\tilde L)$, we construct a smooth 
${d^2 N}{/d\tilde E /d j^2}$ with adaptive kernel density estimation as follows:
\begin{equation}
  \frac{d^2 N}{d\tilde E d j^2}  =  
  \frac{1}{N_{\mathrm{h}}} \sum_{i = 1}^{N_{\mathrm{h}}} 
  \frac{1}{N_{\mathrm{s}, i}} \sum_{k = 1}^{N_{\mathrm{s}, i}}
  \mathcal{S} (\tilde E -\tilde E_{i k}, \sigma_{\tilde E}) 
  \mathcal{S} (j^2 - j^2_{i k}, \sigma_{j^2}),
  \label{eqn:dnej}
\end{equation}
where $N_{\mathrm{h}}$ is the total number of halos in the template sample, 
$N_{\mathrm{s}, i}$ is the total number of satellites in the $i$th halo,
$(\tilde E_{i k}, j^2_{i k})$ refer to the $k$th satellite of the $i$th halo,
and the sums run over every satellite in every halo. In \refeqn{eqn:dnej}, 
$\mathcal{S}$ denotes the Gaussian smoothing kernel, and the kernel sizes
are chosen adaptively as $\sigma_{\tilde  E} = 15 \epsilon_{{\mathrm{sep}}}$ 
and $\sigma_{j^2} = 24 \epsilon_{{\mathrm{sep}}}$, where the difference between
the numerical factors corresponds to that between the standard deviations of
$\tilde  E$ and $j^2$ for satellites in the template sample, and
$\epsilon_{{\mathrm{sep}}}$ is the local average particle separation
as used for smoothed particle hydrodynamics.\footnote{
We estimate the local average separation for a target particle as
$\epsilon_{{\mathrm{sep}}} = d_n \sqrt{\pi / n}$, where $d_n$ is the
distance to the $n$th nearest neighbor (see e.g., \citealt{Price2012}).
We take $n = 50$, but $\epsilon_{{\mathrm{sep}}}$ is not sensitive to $n$
and typically ranges from $\approx 0.02$ to $\approx 0.2$ for the satellites 
in our template sample.}
In the above smoothing procedure, reflecting boundary is used at $\tilde E=0$, 
as well as at $j^2=0$ and 1.

As shown in \reffig{fig:phase_distr}b, the satellites in our template sample
do not cover the upper right corner of the $(\tilde E,\tilde L)$ space that 
lies above the dot-dashed curve and to the right of the solid curve. Because 
this region is physically accessible, we estimate the corresponding 
$\tilde f (\tilde E,\tilde L)$ as follows. We note that the 
solid and dot-dashed curves in \reffig{fig:phase_distr}b start to diverge at
$(\tilde E_{\lim},\tilde L_{\lim})$, where $\tilde E_{\lim}$
corresponds to $\tilde r_{{\mathrm{cir}}} (\tilde E_{\lim})=\tilde r_{\lim}$
and $\tilde L_{\lim}=\tilde L_{\mathrm{cir}}(\tilde E_{\lim})$. For simplicity,
we take $\tilde f (\tilde E, \tilde L) = \tilde f (\tilde E, \tilde L_{\lim})$ 
for the region of $(\tilde E,\tilde L)$ that lies above $\tilde L=\tilde L_{\lim}$
and to the right of the solid curve in \reffig{fig:phase_distr}b.
This estimate is partly based on continuity of $\tilde f (\tilde E, \tilde L)$.
It also follows from the assumption that unbound or nearly unbound
satellites have an isotropic DF with dependence on $\tilde E$ only.
This assumption seems reasonable because high-speed satellites are mainly 
accelerated by the external field, so their direction of motion is largely
unrelated to the host halo. In any case, because most satellites are sufficiently
bound to their host halos, our method of estimating the halo mass is not sensitive 
to the $\tilde f (\tilde E,\tilde L)$ adopted for the region of 
$(\tilde E,\tilde L)$ discussed above.

\section{Validity and robustness of the method} \label{sec:robust}

\begin{figure}[htbp]
  \centering
  \includegraphics[width=0.45\textwidth]{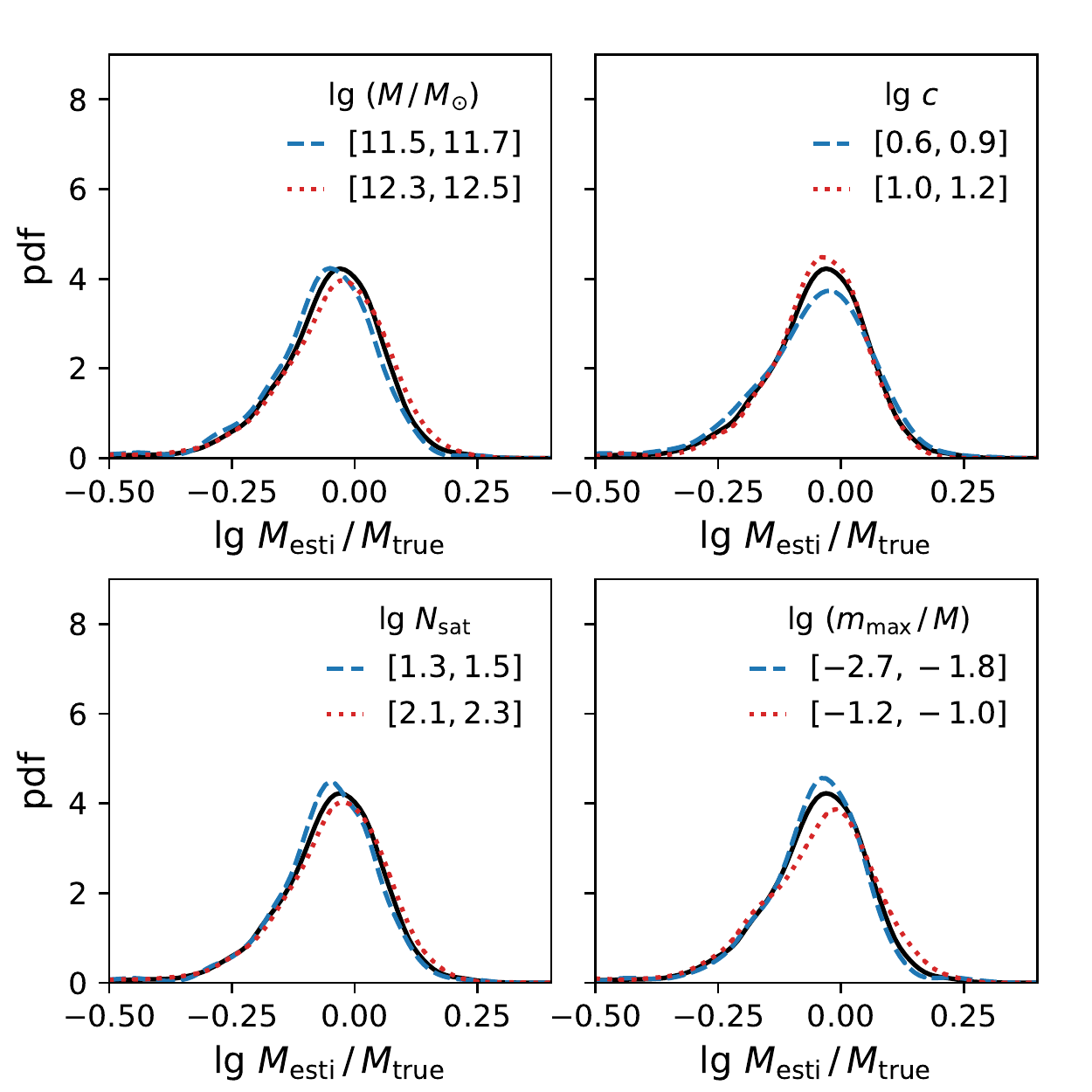}
  \caption{
  Tests of dependence of the precision on various factors.
  \textit{Upper left panel}: Halo mass, $M$. \textit{Upper right panel}: Halo 
  concentration, $c$. \textit{Lower left panel}: Number of satellites,
  $N_\mathrm{sat}$. \textit{Lower right panel}: Ratio of the largest satellite mass at infall,
  $m_\mathrm{max}$, to the halo mass. The black solid curve is the same for all panels
  and shows the distribution of $\lg(M_\mathrm{esti}/M_\mathrm{true})$ for test halos in the 
  whole pertinent sample,
  whereas the red dotted (blue dashed) curves show the distributions for those in
  the top (bottom) 20\% of this sample with the indicated range of the relevant
  factor.
  }
  \label{fig:test_sys}
\end{figure}

Using 20 tracers per halo with a flat pior on $c$, we check the validity and
robustness of our method by applying it to halos in the sample described in 
\refsec{sec:simulation}. Specifically, we test if the precision
of the estimated $\lg M$ depends on the halo mass, the concentration, the richness
of satellites, and the largest satellite mass at infall, respectively. The last 
factor is motivated by the potential influence of the Large Magellanic Cloud on the
mass estimate of the MW halo. We calculate $\lg(M_\mathrm{esti}/M_\mathrm{true})$ 
for test halos in the whole pertinent sample, and for those in the top and bottom 
20\% of this sample, respectively, based on each of the above factors. 
The corresponding distributions are shown in \reffig{fig:test_sys}. It can be seen 
that the precision of the estimated $\lg M$ does not depend on any of the above factors.

\bibliography{MW_MassIII}

\end{document}